\documentstyle[aps,amssymb,pre,psfig]{revtex}
\begin{document}
\bibliographystyle{prsty}
\draft
\tighten
\title{The two--particle problem in	a many--particle system:\\
I. Dynamically screened	ladder approximation}
\author{
Th.	Bornath	and	D. Kremp}
\address{Fachbereich Physik, Universit\"{a}t Rostock,
18051 Rostock, Germany}
\author{M. Schlanges}
\address{Institut f\"ur	Physik,
Ernst--Moritz--Arndt--Universit\"{a}t Greifswald,
17487 Greifswald, Germany
}
\date{\today}
\maketitle
\begin{abstract}
The	two-particle problem within	a nonequilibrium many-particle
system is investigated in the framework	of real-time Green's
functions. Starting	from the dynamically screened ladder approximation
of the nonequlibrium Bethe-Salpeter	equation, a	nonequilibrium Dyson
equation is	given for two-time two-particle	Green's	functions.
Thereby	the	well-known Kadanoff-Baym equations are generalized to the
case of	two-particle functions.	The	two-time structure of the
equations is achieved in an	exact way avoiding the so-called Shindo
approximation. For the case	of thermodynamic equilibrium, the
differences	to former results obtained for the effective
two-particle hamiltonian are discussed.
\end{abstract}
\widetext
\section{Introduction}
This paper is devoted to the kinetic theory	of many-particle
systems	which are able to form bound states. To	be specific, we	will
consider the case of two-particle bound	states,	i.e. the bound states
are	to be thought of, e.g.,	as hydrogen-like atoms or ions.	Our	focus will be
the	derivation of a	kinetic	equation for the distribution function of
the	(possibly excited) bound states. Starting with papers of, for
instance, Waldmann \cite{wald57}, Snider and Lowry
\cite{snider60,snider74}, McLennan and Lagan \cite{l82,lagan84,l89},
Klimontovich and Kremp \cite{kk81},	the	derivation of kinetic
equations has appealed great interest over many	years. Usually (see,
e.g. the review	article	of Klimontovich	et al. \cite{kkk87}) the
two-particle density matrix	is split up	in different parts with
respect	to some	projection operator	which projects onto	the	space of the
bound states. Often	the	states are taken to	be those of	the	isolated
atom. The diagonal matrix elements are considered to be	the
distribution function for the respective bound state. However, in a
dense system, it is	not	clear if the diagonalization of	the	density
matrix with	respect	to the unperturbed two-particle	states is a	good
approximation.

It is well known that many-particle	effects	like dynamical screening,
self-energies or phase space occupation	may	have an	influence on the
two-particle properties. A unique description of these effects within
the	investigation of nonequilibrium	behaviour can be given in the
framework of the real-time Green's functions technique.	For	the
single-particle	functions, this	can	be done	with the Kadanoff-Baym
equations for the correlation functions	$g_a^\gtrless$.	In this	paper
we aim on the derivation of	similar	equations on the two-particle
level.

Some remarks on	the	bound state	problem	in equilibrium seem	to be
useful first. The investigation	of bound states	in dense systems,
which are in the state of thermodynamic	equilibrium, has been the
topic of a lot of papers, see the references cited in the monographs
\cite{ekk76} and \cite{kker86}.	In the framework of	the	Green's
functions method, a	proper starting	point is the so-called
Bethe-Salpeter equation	for	the	two-particle causal	Green's	function
\cite{raja70,zkkkr78,haug78}
\begin{eqnarray}\label{bse}
g_{ab}(12,1'2')	&=&	g_a(1,1') g_b(2,2')\\[1ex]
&&+	i \int d{\bar 1}d{\bar 2}d{\bar{\bar 1}}d{\bar{\bar	2}}\,
 g_a(1,{\bar 1})g_b(2,{\bar	2})	\, K_{ab}({\bar	1}{\bar	2},{\bar{\bar
1}}{\bar{\bar 2}}) \, g_{ab}({\bar{\bar	1}}{\bar{\bar2}},
1'2')\,.\nonumber
\end{eqnarray}
The	kernel of this integral	equation $K_{ab}$ is a four	point
function. In a diagrammatic	language $K_{ab}$ comprises	all
irreducible	diagrams in	the	particle-particle channel. The effective
interaction	$K_{ab}$ has a dynamical character.	This makes the
structure complicate: although one is interested only in the
two-particle Green's function in the particle-particle channel with
$t_1=t_2$ and $t_1'=t_2'$, by the dynamical	kernel $K$ the knowledge
of a Green's function with three times is enforced in the integral
term. In Fourier space (or within the Matsubara	technique),	this is
adequate to	the	problem	that, for the determination	of the
two-particle Green's function dependent	on one frequency, a	more
general	function dependent on two frequencies has to be	known. A way
out	of this	dilemma	has	been tried by applying the Shindo approximation
\cite{shindo} in which the causal quantity with	two	frequencies	is
constructed	from that with one frequency. Then one gets	a closed
equation for the causal	Green's	function. There	are	few	estimations	on
the	range of validity of this approximation. It	is an exact	relation
for	a static interaction $K$. It has been argued that the Shindo
approximation reflects a first order approximation with	respect	to the
retardation	of the effective interaction \cite{kkr83,kker86,schaefer86}.

In order to	evaluate the Bethe-Salpeter	equation, one has, of course,
to use some	approximation for the effective	interaction	$K_{ab}$. In
the	simplest one, which	has	a dynamical	character, $K_{ab}$	is just
the	dynamically	screened potential $V^S_{ab}$. We will come	back to
this point later.

With help of this Bethe-Salpeter equation, some	important questions
concerning the properties of two-particle states in	a plasma could be
discussed. An effective	Schr\"odinger equation was derived which has
some important corrections in comparison with that for an isolated
atom: (i) phase	space occupation factors, (ii) exchange	self-energies
(Hartree-Fock),	(iii) a	dynamically	screened effective potential and
(iv) dynamical single-particle self-energy corrections.	It has been
shown that for localized states	there is to	a large	extent a
compensation between the effects (i) and (iii) on one side and (ii)
and	(iv) on	the	other. It follows that the binding energies	of (at
least the low lying) bound states are not changed considerably in
comparison with	the	isolated atom. In contrast there is	a large	shift
of the continuum edge by the self-energy corrections. This results in a
lowering of	ionization energies	with increasing	plasma density and
leads, at the end, to the well-known Mott effect. An effective wave
equation was solved	numerically	in \cite{seidel95,arndt96},
for	a discussion of	the	result see Kraeft \cite{kraeft99}.

However, the results obtained for the effective	Hamiltonian	have also
some serious shortcomings. There occurs	a division by Pauli-blocking
factors	$1-f_a-f_b$	what causes	spurious pole structures for highly
degenerate systems.	Further, the effective Hamiltonian has static
contributions which	lack a clear physical interpretation (see e.g.
\cite{z88}). These static parts	vanish for a nondegenerate system.

An other approach has been given by	Schuck and co-workers
\cite{dukelskyschuck90,schaeferschuck99}. They postulate that Dyson
equations exist	for	two-time causal	and	retarded Green's functions,
respectively. Expressions for the self-energy operator (also called
mass operator or effective Hamiltonian)	are	derived	by comparison with
the	respective equations of	motion.	It remains also	unclear	in this
approach what approximation	(if	any) is	connected with the assumption
that such two-time Dyson equations for the investigated	functions
and	the	inverse	of those functions,	respectively, do exist.	Also in
this approach, there occours the problem of	the	division by
Pauli-blocking factors.

Looking	at the complicate structure	of the Bethe-Salpeter equation,
there arises the question if a general formulation in terms	of a
two-time Dyson equation	is possible. The approach using	the	Shindo
approximation is {\it per se} an approximative one.	Schuck et al. do
not	consider the Bethe-Salpeter	equation at	all	but	postulate just
this two-time structure.
There are some strong arguments	in favour of the possibility that
exact equations	exist for two-time Green's functions which contain
solely two-time	quantities (the	inner structure	of a self-energy
operator could be complicate, however).
In a chemical picture, for instance, in	which
bound states are interpreted as	a new species, one would expect
equations for correlation functions	$g^\gtrless_{ab}(t,t')$
similar	to the Kadanoff-Baym equations in the single-particle case
with some two-particle self-energy functions. The spectral
information, e.g. binding energies,	damping	etc., should follow	from
the	retarded two-particle Green's function $g^R_{ab}(t,t')$. The
correlation	function $g^<_{ab}(t,t')$ is for $t=t'$	just the
two-particle density matrix.

In nonequilibrium one can derive an	equation of	the	same structure
like in	Eq.~(\ref{bse}), however, the time integrations	have to	be
performed then on the Keldysh contour. The functions involved in the
integral term consist of many correlation functions	because	they are
dependent on three times at	least. Sch\"afer et	al.	\cite{schaefer88}
considered the dynamically screened	ladder approximation for the
polarizability in a	semiconductor within the Keldysh formalism.
They gave a	formulation	for	functions depending	on three times or --
after Fourier transformation --	on two frequencies.	At the end,
however, they used the Shindo approximation	for	these two-frequency
quantities in order	to get kinetic equations for single-frequency
functions.

There were attempts	to generalize
the	Shindo approximation to	functions in the time domain
\cite{kks98}. It was also tried, \cite{bornath98}, to generalize the
approach of	Schuck et al. within the nonequilibrium	real-time Green's
functions method postulating a Dyson equation for the retarded
function $g_{ab}^R(t,t')=\Theta(t-t')(g_{ab}^>-g_{ab}^<)$. In both
approaches similar results were	achieved. The equilibrium results
could be reproduced. Thus the same shortcomings	arise for degenerate
systems.

We will	present	here a new approach	\cite{bsk99} to	this problem
within the real-time Green's functions method. Thus	we are able	to
describe nonequlibrium systems.	Results	for	thermodynamic equlibium
will appear	as special case	of the more	general	equations. In this
first part the nonequilibrium Bethe-Salpeter equation is considered	in
a concrete approximation, the so-called	dynamically	screened ladder
equation. This is the simplest approximation in	which the effective
interaction	is of dynamical	nature.	This will enable us	to identify
the	underlying algebraic structures	and	to keep	the	equations as
simple as possible.	The	general	scheme will	be investigated	in a
subsequent paper \cite{bor99}.

The	structure of this paper	is as follows. In Sec.~\ref{spq} the
scheme of the real-time	Green's	function method	for	single-particle
Green's	functions is summarized. The difficulties of the
Bethe-Salpeter equation	are	discussed in Sec.~\ref{besal}.

The	dynamically	screened ladder	approximation is considered	in
Sec.~\ref{dsla}. The Bethe-Salpeter	equation is	written	down in	this
approximation for two-time functions on	the	Keldysh	double time
contour. The first terms of	the	pertubation	expansion with respect to
the	dynamically	screened potential are evaluated. The two-time
structure will be achieved by applying the semi-group property of the
ideal propagators. After that, the algebraic structures	turn out to	be
similar	to those of	the	non-equilibrium	Dyson-Keldysh equation in the
single-particle	case and the two-particle self-energy functions	can	be
identified by comparison. The thermodynamic	equilibrium	case is
considered in Sec.~\ref{tde}. The structure	of the two-particle
self-energy	which can be understood	as an effective	Hamiltonian
is discussed. The results are compared with	the	former ones
\cite{zkkkr78,kkr83,kker86,z88}.
It will	turn out that only in the nondegenerate	case and in	the	static
limit, one is led to the same results.

%%%%%%%%%%%%%%%%%%%%%%%%%%%%%%%%%%%%%%%%%%%%%%%%%%%%%%%%%%%%%%%%%%%%%
\section{Single-particle quantities}\label{spq}
%%%%%%%%%%%%%%%%%%%%%%%%%%%%%%%%%%%%%%%%%%%%%%%%%%%%%%%%%%%%%%%%%%%%%
Let	us summarize shortly the scheme	of real-time Green's function
technique in the single-particle case. The equations are given on a
double-time	contour, on	the	so-called Keldysh contour
\cite{keldysh64,dubois67}.
Objects	of the algebra are matrices	of causal and anticausal
Green's	functions, $g_a$ and $\overline{g}_a$, and the correlation
functions $g_a^\gtrless$ defined below
\begin{eqnarray}\label{definition}
g_{a}({\bf r}_1t,{\bf r}_1't')&=&\Theta(t-t')
g_{a}^>({\bf r}_1t,{\bf	r}_1't')
+\Theta(t'-t) g_{a}^<({\bf r}_1t,{\bf r}_1't')\\
&=&\Theta(t-t')\frac{1}{i}\langle\Psi_a({\bf r}_1,t)
\Psi_a^\dagger({\bf	r}_1',t') \rangle
+
\Theta(t'-t)(\pm)\frac{1}{i}\langle
\Psi_a^\dagger({\bf	r}_1',t')
\Psi_a({\bf	r}_1,t)	\rangle\,
\,,
\nonumber
\\[2ex]
\overline{g}_{a}({\bf r}_1t,{\bf r}_1't')&=&\Theta(t-t')
g_{a}^<({\bf r}_1t,{\bf	r}_1't')
+\Theta(t'-t) g_{a}^>({\bf r}_1t,{\bf r}_1't')\,.
\nonumber
\end{eqnarray}
Here $a$ is	the	species	index, $\Psi_a^\dagger$	and	$\Psi_a$ are
creation and annihilation operators	of second quantization with
the	commutation	relations
\begin{eqnarray}\label{commu}
[\Psi_a({\bf r},t),\Psi_b^\dagger({\bf r}',t)]_{\mp}&=&
\delta_{ab}\delta({\bf r}-{\bf r}')\\ \nonumber
[ \Psi_a({\bf r},t),\Psi_b ({\bf r}',t)]_{\mp}&=&0\\ \nonumber
[ \Psi^\dagger_a({\bf r},t),\Psi^\dagger_b ({\bf r}',t)]_{\mp}&=&0
 \,	.
\end{eqnarray}
The	upper sign (meaning	the	commutator)	holds for Bosons and the lower
one	(anticommutator) for Fermions. Spin	is not written explicitely here.

One	can	see	that these elements	are	not	all	independent.
It turns out that the equations	get	a more convenient structure	if one
introduces two other quantities, $g_a^R$ and $g_a^A$, defined by
\begin{eqnarray}\label{g1-R}
g_a^R&=&\Theta(t-t')\frac{1}{i}\langle\Psi_a(r_1,t)
\Psi_a^\dagger(r_1',t')\mp \Psi_a^\dagger(r_1',t')\Psi_a(r_1,t)
\rangle\\
&=&\Theta(t-t')\Big[g_a^>(t,t')-g_a^<(t,t')\Big]
\,,
\nonumber
\\
g_a^A&=&-\Theta(t'-t)\frac{1}{i}\langle\Psi_a(r_1,t)
\Psi_a^\dagger(r_1',t')\mp \Psi_a^\dagger(r_1',t')\Psi_a(r_1,t)
\rangle\\
&=&\Theta(t'-t)\Big[g_a^<(t,t')-g_a^>(t,t')\Big]\nonumber
\,.
\end{eqnarray}
It follows that
\begin{eqnarray}\label{retardGF}
g_a^R&=&g_a-g_a^<=g_a^>-\overline{g}_a\,, \\
g_a^A&=&g_a-g_a^>=g_a^<-\overline{g}_a \,.\nonumber
\end{eqnarray}

\begin{figure}[h]
\centerline{\psfig{figure=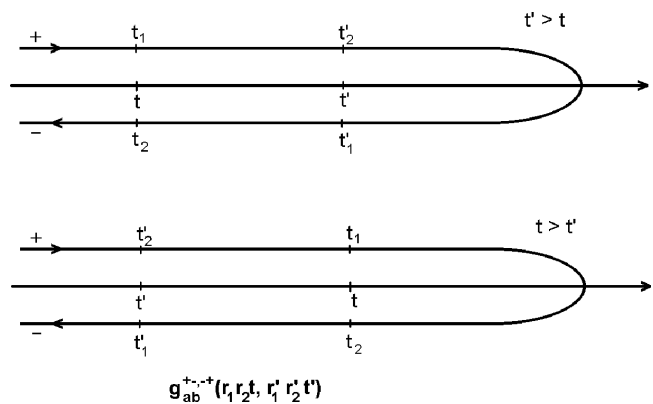,width=8.5cm,angle=0}}
\vspace*{0.5cm}
\caption[]{
Time ordering on the Keldysh double	time contour for the function
$g_{ab}^{+-,-+}$. The ordering is causal on	the	upper branch and
anticausal on the lower	branch.
}\label{contour}
\end{figure}

The	nonequlilibrium	Dyson equation on  the Keldysh contour reads
\begin{eqnarray}\label{dyson-cont}
{\underline	g}_a(1,1')&=&{\underline g}_{a,0}(1,1')
+\int\limits_{\cal C} d{\bar 1}	d{\bar{\bar	1}}\,
{\underline	g}_{a,0}(1,{\bar 1})\,
{\underline	\Sigma}_a({\bar	1},{\bar{\bar 1}})\, {\underline
g}_a({\bar{\bar	1}},1')\,,
\end{eqnarray}
with $1={\bf r}_1,t_1$ etc., and ${\underline g}_{a,0}$	being the
ideal functions	and	$\Sigma_a$ the self-energy.	The	time integrations
are	performed on the Keldysh contour, see Fig.~\ref{contour}. The
underlined quantities are causal ones for both times on	the	upper
branch of the contour, $g_a=g_a^{++}$, and anticausal ones for both
times on the lower branch, ${\overline g}_a=g_a^{--}$. If the first
time is	on the upper branch	and	the	second one on the lower, one gets
$g_a^<=g_a^{+-}$. Fixing the first time	on the lower and the second
time on	the	upper branch gives $g_a^>=g_a^{-+}$. Working on	the
Keldysh	contour	has	the	advantage that well-developed schemes of
functional derivatives and diagrammatic	techniques known from
equilibrium	\cite{kb62,blaizot86} can easily be	generalized	to
nonequilibrium situations, see,	e.g., \cite{dubois67,d84,d90,boter90}.

Eq.~(\ref{dyson-cont}) is equivalent to	a set of four
equations which	are, however, not all independent. Therefore, it is
often more convenient to work with the following form of the
nonequlibrium Dyson	equation for the correlation function $g_a^<$
\begin{eqnarray}\label{dyson1}
g_a^<(t,t')&=&g^<_{a,0}(t,t')+\int\limits_{t_0}^\infty
dt_1\,\int\limits_{t_0}^\infty dt_2\, \Big[g^<_{a,0}(t,t_1)\,
\Sigma_a^A(t_1,t_2)\, g_a^A(t_2,t')\\
&& + g^R_{a,0}(t,t_1)\,\Sigma_a^<(t_1,t_2)\,
g_a^A(t_2,t') +	g^R_{a,0}(t,t_1)\,\Sigma_a^R(t_1,t_2)\,
g_a^<(t_2,t') \Big]\,, \nonumber
\end{eqnarray}
which has to be	supplemented by	an equation	for	$g_a^{R/A}$
\begin{eqnarray}\label{dyson1R}
g_a^R(t,t')&=&g^R_{a,\,0}(t,t')+\int\limits_{t_0}^\infty
dt_1\,\int\limits_{t_0}^\infty dt_2\,g^R_{a,\,0}(t,t_1)\,
\Sigma_a^R(t_1,t_2)\,g_a^R(t_2,t')
\,.
\end{eqnarray}
Here only the functions's dependence on	the	times was written
explicitely	in order to	save space.	Often the initial time is
considered in the limit	$t_0\longrightarrow	-\infty$. The quantity
$g_a^A$	is connected with $g_a^R$ by hermitean conjugation
\begin{equation}
g_{a}^{A }({\bf	r} t, {\bf r}^\prime
t^\prime)=\left[g_{a}^{R}({\bf r}^\prime t^\prime, {\bf	r}
t)\right]^\ast \,.
\end{equation}
An other form of Eq.~(\ref{dyson1})	one	can	find is	\cite{d84}
\begin{eqnarray}\label{dyson11}
g_a^<(t,t')&=&g_a^R(t,t_0)\,g^<_a(t_0,t_0)\,g^A_{a}(t_0,t')
+\int\limits_{t_0}^\infty dt_1\,\int\limits_{t_0}^\infty dt_2\,
g^R_a(t,t_1)\,\Sigma_a^<(t_1,t_2)\,	g_a^A(t_2,t')\,.
\end{eqnarray}

The	corresponding differential equations read
\begin{eqnarray}\label{kineq_1}
\left[i\frac{\partial}{\partial	t}+\frac{\nabla^2}{2m_a}
\right]\,g_{a}^<(t,t')&=&\int d{\bar t}\,\Big[
\Sigma_{a}^R(t,{\bar t})\,g_{a}^<({\bar	t},t') +
\Sigma_{a}^<(t,{\bar t})\,G_{a}^A({\bar	t},t')\Big]\,,\\
\left[i\frac{\partial}{\partial	t}+\frac{\nabla^2}{2m_a}
\right]\,g_{a}^R(t,t')&=&\delta(t-t')+\int d{\bar t}\,
\Sigma_{a}^R(t,{\bar t})\,g_{a}^R({\bar	t},t')\,.
\end{eqnarray}
These are the well-known generalized kinetic equations (Kadanoff-Baym
equations) for the single-particle functions.

In the following chapters
the	one-particle self-energy will be needed	in a special
approximation which	is called $V^S$-approximation. Here	one	has
\begin{eqnarray}\label{sigma-one}
 {\underline \Sigma}_a(1,1')&=&{\underline \Sigma}_a^{H}
 +
%{\underline {\bar \Sigma}}_a(1,1')\,,\\
%{\underline {\bar \Sigma}}_a(1,1')&=&
i{\underline V}^{S\,}_{aa}(1,1')\,{\underline g}_a(1,1')
\,,
\end{eqnarray}
with $\Sigma^H$	being the Hartree self-energy.
As for the Green's functions, cf. Eqs.~(\ref{definition},\ref{retardGF}),
there is a set of functions	describing
the	dynamically	screened interaction ${\underline V}_{ab}^S$
\begin{eqnarray}\label{vs}
V^S_{ab}(t,t')&=&V_{ab}\,\delta(t-t')+\Theta(t-t')
V_{ab}^{S\,>}(t,t')+\Theta(t'-t) V_{ab}^{S\,<}(t,t')
\,,
\\[1ex]
\overline{V}^S_{ab}(t,t')&=&-V_{ab}\,\delta(t-t')+\Theta(t-t')
V_{ab}^{S\,<}(t,t')+\Theta(t'-t) V_{ab}^{S\,>}(t,t')
\,,
\nonumber
\\[1ex]
V_{ab}^{S\,R}(t,t')&=&V_{ab}\,\delta(t-t')+\Theta(t-t')	\Big[
V_{ab}^{S\,>}(t,t')- V_{ab}^{S\,<}(t,t')\Big]
\,,
\nonumber
\\[1ex]
V_{ab}^{S\,A}(t,t')&=&V_{ab}\,\delta(t-t')+\Theta(t'-t)	\Big[
V_{ab}^{S\,<}(t,t')- V_{ab}^{S\,>}(t,t')\Big]
\,.
\nonumber
\end{eqnarray}
Here the correlation functions are defined by
\begin{eqnarray}
V_{ab}^{S\,\gtrless}(t,t')=\sum_{c,d} V_{ac} L^\gtrless_{cd}(t,t')
V_{db}
\,,
\end{eqnarray}
where $L^\gtrless$ are the correlation functions of	density
fluctuations
\begin{eqnarray}
iL_{ab}^>({\bf r}_1	t_1, {\bf r}_2 t_2)&=&\langle\delta
\hat\rho_a({\bf	r}_1,t_1)\delta	\hat\rho_b({\bf	r}_2,t_2)\rangle
\,,
\\
iL_{ab}^<({\bf r}_1	t_1, {\bf r}_2 t_2)&=&\langle\delta
\hat\rho_b({\bf	r}_2,t_2)\delta	\hat\rho_a({\bf	r}_1,t_1)\rangle
\,,
\nonumber
\end{eqnarray}
with $\delta \hat
\rho_a({\bf	r},t)=\Psi_a^\dagger({\bf r},t)\Psi_a({\bf r},t)
-\langle\Psi_a^\dagger({\bf	r},t)\Psi_a({\bf r},t)\rangle$.\\
It follows that, e.g.,	$V_{ab}^{S\,<}(1,1')=V_{ba}^{S\,>}(1',1)$ and
$V_{ab}^{S}(1,1')=V_{ba}^{S}(1',1)$, but
$V_{ab}^{S\,R}(1,1')=V_{ba}^{S\,A}(1',1)$.

%%%%%%%%%%%%%%%%%%%%%%%%%%%%%%%%%%%%%%%%%%%%%%%%%%%%%%%%%%%%%%%%%%%%
\section{Bethe-Salpeter	equation}\label{besal}
%%%%%%%%%%%%%%%%%%%%%%%%%%%%%%%%%%%%%%%%%%%%%%%%%%%%%%%%%%%%%%%%%%%%
The	two-particle Green's function is determined	by
the	so-called Bethe-Salpeter equation
\begin{eqnarray}\label{bse-keldysh}
g_{ab}(12,1'2')	&=&	g_a(1,1') g_b(2,2')\\[1ex]
&&+	i \int d{\bar 1}d{\bar 2}d{\bar{\bar 1}}d{\bar{\bar	2}}\,
 g_a(1,{\bar 1})g_b(2,{\bar	2})	\, K_{ab}({\bar	1}{\bar	2},{\bar{\bar
1}}{\bar{\bar 2}}) \, g_{ab}({\bar{\bar	1}}{\bar{\bar2}},1'2')
\,,\nonumber
\end{eqnarray}
in which by	introduction of	the	effective interaction kernel $K_{ab}$
formally a closed equation is achieved for the four-point function.
Here, the kernel $K_{ab}$ is the sum of	all	diagrams irreducible with
respect	to a cutting of	two	single-particle	lines. The Bethe-Salpeter
equation can be	understood to hold in various contexts:	for	$T=0$, for
the	imaginary-time equilibrium Green's functions in	the	Matsubara
technique, or for the real-time	Green's	functions on the Keldysh
contour.

The	properties of a	pair of	particles should follow	from this equation
in the so-called particle-particle channel.	If one considers the
causal two-particle	Green's	function in	this channel ($t_1=t_2=t;
t_1'=t_2'=t'$),	one	has
\begin{eqnarray}\label{g-causal}
i^2	g_{ab}(r_1r_2t,r_1'r_2't')&=&\theta(t-t')\Big\langle
\Psi_a({\bf	r}_1,t)\Psi_b({\bf r}_2,t)
\Psi_b^\dagger({\bf	r}_2',t')\Psi_a^\dagger({\bf r}_1',t')
 \Big\rangle \\
& +&
\theta(t'-t)\Big\langle
\Psi_a^\dagger({\bf	r}_1',t')\Psi_b^\dagger({\bf r}_2',t')
\Psi_b({\bf	r}_2,t)\Psi_a({\bf r}_1,t) \Big\rangle\nonumber\\
&=&\theta(t-t')i^2g_{ab}^>+\theta(t'-t)i^2g_{ab}^<
\nonumber\,.
\end{eqnarray}
On the right hand side of Eq.~(\ref{bse-keldysh}), however,	there
occurs a function depending	on three times ${\bar t}_1$, ${\bar	t}_2$,
and	$t'$ what is enforced by the dynamical character of	$K_{ab}$.
This Green's function consists of six different	correlation	functions
\begin{eqnarray}\label{dreizeit}
i^2\,g_{ab}({\bar r}_1{\bar	t}_1{\bar r}_2{\bar	t}_2;r_1'r_2't')
&=&\theta({\bar	t}_1-{\bar t}_2)\theta({\bar t}_2-t')
\Big\langle\Psi_a({\bar	{\bf r}}_1,{\bar t}_1)
\Psi_b({\bar {\bf r}}_2,{\bar t}_2)
\Psi_b^\dagger({\bf	r}_2',t')
\Psi_a^\dagger({\bf	r}_1',t')
 \Big\rangle\\
&\pm&
\theta({\bar t}_2-{\bar	t}_1)\theta({\bar t}_1-t')
\Big\langle
\Psi_b({\bar {\bf r}}_2,{\bar t}_2)
\Psi_a({\bar {\bf r}}_1,{\bar t}_1)
\Psi_b^\dagger({\bf	r}_2',t')
\Psi_a^\dagger({\bf	r}_1',t')
 \Big\rangle  \nonumber\\
&+&
\theta({\bar t}_1-t')\theta(t'-{\bar t}_2)
\Big\langle
\Psi_a({\bar {\bf r}}_1,{\bar t}_1)
\Psi_b^\dagger({\bf	r}_2',t')
\Psi_a^\dagger({\bf	r}_1',t')
\Psi_b({\bar {\bf r}}_2,{\bar t}_2)
 \Big\rangle\nonumber\\
& +&
\theta({\bar t}_2-t')\theta(t'-{\bar t}_1)
\Big\langle
\Psi_b({\bar {\bf r}}_2,{\bar t}_2)
\Psi_a^\dagger({\bf	r}_1',t')
\Psi_b^\dagger({\bf	r}_2',t')
\Psi_a({\bar {\bf r}}_1,{\bar t}_1)
 \Big\rangle\nonumber\\
&\pm&
\theta(t'-{\bar	t}_1)\theta({\bar t}_1-{\bar t}_2)
\Big\langle
\Psi_a^\dagger({\bf	r}_1',t')
\Psi_b^\dagger({\bf	r}_2',t')
\Psi_a({\bar {\bf r}}_1,{\bar t}_1)
\Psi_b({\bar {\bf r}}_2,{\bar t}_2)
 \Big\rangle\nonumber\\
& +&
\theta(t'-{\bar	t}_2)\theta({\bar t}_2-{\bar t}_1)
\Big\langle
\Psi_a^\dagger({\bf	r}_1',t')
\Psi_b^\dagger({\bf	r}_2',t')
\Psi_b({\bar {\bf r}}_2,{\bar t}_2)
\Psi_a({\bar {\bf r}}_1,{\bar t}_1)
 \Big\rangle \nonumber
\,.
\end{eqnarray}
Only a static interaction in (\ref{bse-keldysh}) would enforce ${\bar
t}_1={\bar t}_2$, and the function would turn into the two-time	causal
one	(\ref{g-causal}).

In principle, one could	of course try to solve the Bethe-Salpeter
equation for a function	depending on three times (equivalent to	a
function depending on two frequencies) and then	extract	from this the
information	one	is interested in. This,	however	seems to be	to
complicated. In	a number of	papers it was tried	therefore to work with
equations which	involve	exclusively	two-time functions.	This was
achieved in	two	ways. The first	approach \cite{zkkkr78,haug78}
uses within	the	Matsubara
technique the so-called	Shindo approximation in	which the
two-frequency function is constructed from the single-frequency	causal
Green's	function. This is possible in an exact way for a statical
interaction	and	therefore it was argued	that for arbitrary $K_{ab}$
this would be correct in first order with respect to the retardation.
In the other approach \cite{dukelskyschuck90}, a closed	equation for
the	causal two-time	or single-freqency,	respectively, function is
postulated.	The	effective hamiltonian (two-particle	self-energy) is
determined then	by comparison with equations of	motion.

In any case	there are closed equations for combinations	of the
correlation	functions $g_{ab}^\gtrless$	defined	in (\ref{g-causal}).
But	this means that	the	information	contained in the correlation
functions
$\langle\Psi_a\Psi_b^\dagger
\Psi_a^\dagger\Psi_b\rangle$
and
$\langle\Psi_b\Psi_a^\dagger
\Psi_b^\dagger\Psi_a\rangle$
(third and fourth term in Eq.~\ref{dreizeit}) is neglected.
Therefore such a closed	equation for the causal	two-time Green's
function can exist only	in an approximate way.

In the next	section	we will	show that closed equations exist for
other combinations of correlation functions.

\section{Dynamically screened ladder approximation}\label{dsla}
In analogy to the single-particle case,	one	is interested to get
information	on the statistical properties, carried by the two-particle
density	matrix,	as well	as information on the two-particle dynamics,
the	spectral information. Below	we will	see	that it	is not a trivial
question to	say	which quantity carries this	information.

One	of the quantities of interest is the following two-time
correlation	function
\begin{eqnarray}\label{gsmaller}
g^<_{ab}({\bf r}_1 {\bf	r}_2 t,	{\bf r}_1' {\bf	r}_2' t'\,)=
\frac{1}{i^2}\Big\langle
\Psi_a^\dagger({\bf	r}_1',t')\Psi_b^\dagger({\bf r}_2',t')
\Psi_b({\bf	r}_2,t)\Psi_a({\bf r}_1,t) \Big\rangle\,.
\end{eqnarray}
In the case	$t'=t$,	the	quantity $i^2\,g_{ab}^<$ is
just the two-particle density matrix
$\rho_{ab}({\bf	r}_1 {\bf r}_2 {\bf	r}_1' {\bf r}_2', t\,)$.

We will	use	in the following the real-time Green's function	technique
in the Keldysh formulation.	In the present section the Bethe-Salpeter
equation will be considered	in a concrete approximation: for the
effective interaction kernel $K_{ab}$, the dynamically screened
potential $V^S_{ab}$ is	taken. Then	one	has	on the Keldysh contour
\begin{equation}\label{bse-vs}
{\underline	g}_{ab}(t_1t_2,t_1't_2')={\underline g}_a(t_1,t_1')\,
{\underline	g}_b(t_2,t_2')+\int_{\cal C} d{\bar	t}_1\,d{\bar t}_2\,
{\underline	g}_a(t_1,{\bar t}_1)\,{\underline g}_b(t_2,{\bar t}_2)\,
i\underline{V}_{ab}^S({\bar	t}_1,{\bar t}_2)
{\underline	g}_{ab}({\bar t}_1 {\bar t}_2,t_1't_2')\,.
\end{equation}
Iteration of this integral equation	leads to ladder-type terms.

This Bethe-Salpeter	equation (\ref{bse-vs})	will be	considered in the
following in the
special	case $t_1=t_2=t$ and $t_1'=t_2'=t'$.
%\newpage
\subsection{Two-particle Green's functions depending on	two	times}
Although the physical times	$t_1$ and $t_2$	as well	as $t_1'$ and
$t_2'$ are equal, there	are	still
16 possibilities to	fix	the	times $t_1,	t_2, t_1',$	and	$t_2'$ on the
upper and lower	branches of	the	Keldysh	contour.
Fixing the times $t_1$
and	$t_2$ on the upper and $t_1'$
and	$t_2'$ on the lower	branch of the Keldysh contour, one gets
an equation	for	the	correlation	function $g_{ab}^{++,--}=g_{ab}^<$
defined	in Eq.~(\ref{gsmaller}).
Below there	are	given three	other important	cases.
\begin{eqnarray}
\label{g+--+}
g_{ab}^{+-,-+}({\bf	r}_1 {\bf r}_2 t, {\bf r}_1' {\bf r}_2'	t'\,)
&=&\Theta(t-t')\frac{1}{i^2}\Big\langle
\Psi_a^\dagger({\bf	r}_1',t')\Psi_b({\bf r}_2,t)
\Psi_a({\bf	r}_1,t)\Psi_b^\dagger({\bf r}_2',t') \Big\rangle\\
&&+\Theta(t'-t)\frac{1}{i^2}\Big\langle
\Psi_b({\bf	r}_2,t)\Psi_a^\dagger({\bf r}_1',t')
\Psi_b^\dagger({\bf	r}_2',t')\Psi_a({\bf r}_1,t) \Big\rangle
\nonumber\,,\\
\label{g-++-}
g_{ab}^{-+,+-}({\bf	r}_1 {\bf r}_2 t, {\bf r}_1' {\bf r}_2'	t'\,)
&=&\Theta(t-t')\frac{1}{i^2}\Big\langle
\Psi_b^\dagger({\bf	r}_2',t')\Psi_a({\bf r}_1,t)
\Psi_b({\bf	r}_2,t)\Psi_a^\dagger({\bf r}_1',t')
\Big\rangle\\
&&+\Theta(t'-t)\frac{1}{i^2}\Big\langle
\Psi_a({\bf	r}_1,t)\Psi_b^\dagger({\bf r}_2',t')
\Psi_a^\dagger({\bf	r}_1',t')\Psi_b({\bf r}_2,t) \Big\rangle
\nonumber\,,\\
\label{g--++}
g_{ab}^{--,++}({\bf	r}_1 {\bf r}_2 t, {\bf r}_1' {\bf r}_2'	t'\,)
&=&\frac{1}{i^2}\Big\langle
\Psi_a({\bf	r}_1,t)\Psi_b({\bf r}_2,t)
\Psi_b^\dagger({\bf	r}_2',t')\Psi_a^\dagger({\bf r}_1',t')
\Big\rangle\,.
\end{eqnarray}
To give	an example,	the	time-ordering on the Keldysh contour is	shown
for	$g_{ab}^{+-,-+}$ in	Fig.~\ref{contour}.
All	other functions	$g_{ab}^{\alpha	\beta,\gamma \delta}$ with the
greek indices equal	to ``$+$'' or ``$-$'' can be expressed in terms	of
the	six	correlation	functions involved in Eqs.~(\ref{gsmaller})	and
(\ref{g+--+}-\ref{g--++}).

We define the following	retarded and advanced quantities
\begin{eqnarray}\label{retard-G}
G_{ab}^R({\bf r}_1 {\bf	r}_2 t,	{\bf r}_1' {\bf	r}_2' t'\,)
&\equiv& \Theta(t-t')i\Big(g_{ab}^{++,--}-g_{ab}^{+-,-+}-g_{ab}^{-+,+-}
+g_{ab}^{--,++}\Big)\\
&=&\Theta(t-t')\frac{1}{i}\Big\langle
\Big[\Psi_a^\dagger({\bf r}_1',t'),
\Big[\Psi_b^\dagger({\bf r}_2',t'),
\Psi_b({\bf	r}_2,t)\Psi_a({\bf r}_1,t)
\Big]_{-}\,
\Big]_{\mp}\,
\Big\rangle\nonumber
\,,
\\[1ex]
G_{ab}^A({\bf r}_1 {\bf	r}_2 t,	{\bf r}_1' {\bf	r}_2' t'\,)
&\equiv	&\Theta(t'-t)(-i)\Big(g_{ab}^{++,--}-g_{ab}^{+-,-+}-g_{ab}^{-+,+-}
+g_{ab}^{--,++}\Big)
\label{advanc-G}
\\
&=&\Theta(t'-t)\frac{1}{-i}\Big\langle
\Big[
\Psi_a({\bf	r}_1,t)
,
\Big[
\Psi_b({\bf	r}_2,t)
,
\Psi_b^\dagger({\bf	r}_2',t')
\Psi_a^\dagger({\bf	r}_1',t')
\Big]_{-} \,
\Big]_{\mp}	\,
\Big\rangle
\,.
\nonumber
\end{eqnarray}
In order to	achieve	the	nested commutator structures, it was used that
operators with equal times can be interchanged according to
(\ref{commu}). Interestingly enough	these nested structures	were found
also by	Rajagopal and Majumdar in their	analysis of	double dispersion
relations for the two-frequency	causal Matsubara Green's function (Appendix	II
of \cite{raja70}). We show in Appendix \ref{analyt}	as an example
how	the	function $G_{ab}^{A}$ is connected in thermodynamic
equilibrium	with the analytic continuation of the two-frequency
Matsubara Green's function.

The	functions $G_{ab}^{R/A}$ have the following	properties:\\
(i)	They are
connected by hermitean conjugation
\begin{equation}
G_{ab}^R({\bf r}_1{\bf r}_2	t, {\bf	r}_1'{\bf r}_2't')=
\Big[G_{ab}^A({\bf r}_1'{\bf r}_2' t', {\bf	r}_1{\bf
r}_2t)\Big]^\dagger
\,.
\end{equation}
(ii) Both functions	have the property of crossing symmetry,	i.e.
\begin{equation}
G_{ab}^{R/A}({\bf r}_1{\bf r}_2	t, {\bf	r}_1'{\bf r}_2't')=
G_{ba}^{R/A}({\bf r}_2{\bf r}_1	t, {\bf	r}_2'{\bf
r}_1't')
\,.
\end{equation}
(iii) The inhomogeneity	in the equations of	motion for these functions
consists of	$\delta$-functions only	(without Pauli-blocking	terms).
This is	easy to	see	from Eqs.~(\ref{retard-G},\ref{advanc-G}).
Derivation of the Heaviside-function gives a $\delta$-function
$\delta(t-t')$,	and	the	commutation	relations for the field	operators
lead to	$[\delta({\bf r}_1-{\bf	r}_1')\delta({\bf r}_2-{\bf	r}_2')\pm
\delta_{ab}\delta({\bf r}_1-{\bf r}_2')\delta({\bf r}_2-{\bf
r}_1')]$.\\
(iv) For vanishing interaction between the particles $a$ and $b$, the
correlation	functions in Eqs.~(\ref{retard-G},\ref{advanc-G}) can be
contracted into	products of	single-particle	correlation	functions
$g^\gtrless$, and one gets
\begin{eqnarray}
G_{ab}^R({\bf r}_1 {\bf	r}_2 t,	{\bf r}_1' {\bf	r}_2'
t'\,)&=&i\,g_{a}^R({\bf	r}_1 t,	{\bf r}_1't')
g_{b}^R({\bf r}_2 t, {\bf r}_2't')
\pm	\delta_{ab}\, i\,g_{a}^R({\bf r}_1 t, {\bf r}_2't')
g_{b}^R({\bf r}_2 t, {\bf r}_1't')\,,
\end{eqnarray}
with the retarded single-particle function $g_a^R$ defined by
Eq.~(\ref{g1-R}).\\
(v)	The	difference of the retarded and the advanced	functions defines
a spectral function
\begin{eqnarray}
A_{ab}({\bf	r}_1 {\bf r}_2 t, {\bf r}_1' {\bf r}_2'	t'\,)&=&
iG_{ab}^R({\bf r}_1	{\bf r}_2 t, {\bf r}_1'	{\bf r}_2' t'\,)
-iG_{ab}^A({\bf	r}_1 {\bf r}_2 t, {\bf r}_1' {\bf r}_2'	t'\,)\\
&=&i^2\Big[g_{ab}^{++,--}-g_{ab}^{+-,-+}-g_{ab}^{-+,+-}
+g_{ab}^{--,++}\Big]\nonumber
\,,
\end{eqnarray}
which gives	in the case	of equal times ($t=t'$)
\begin{eqnarray}
A_{ab}({\bf	r}_1 {\bf r}_2 t, {\bf r}_1' {\bf r}_2'	t\,)=
\delta({\bf	r}_1-{\bf r}_1')\delta({\bf	r}_2-{\bf r}_2')\pm
\delta_{ab}\delta({\bf r}_1-{\bf r}_2')\delta({\bf r}_2-{\bf
r}_1')
\,.
\end{eqnarray}
%%%%%%%%%%%%%%%%%%%%%%%%%%%%%%%%%%%%%%%%%%%%%%%%%%%%%%%%%%%%%%%%%%%%%
\subsection{Transformation of the Bethe-Salpeter equation into a Dyson
equation}
%%%%%%%%%%%%%%%%%%%%%%%%%%%%%%%%%%%%%%%%%%%%%%%%%%%%%%%%%%%%%%%%%%%%%
In this	subsection the dynamically screened	ladder equation	as a
special	approximation of the BSE is	transformed	into a Dyson equation
in which the occuring two-particle Green's functions and two-particle
self-energy	functions are dependent	on two times only. For this
purpose, the perturbation expansion	of the Bethe-Salpeter equation
(\ref{bse-vs}) is considered in	the	diagrammatic form shown	in
Fig.~\ref{diagram}.	It is analyzed first for $g_{ab}^<$; details are
presented for different	orders of $V_{ab}^S$ in	Appendix
\ref{orde}.

We search for (and,	indeed,	find) the following	structures
[cf. the corresponding equations for the single-particle functions,
Eqs. \ref{dyson1} and \ref{dyson1R}]
\begin{eqnarray}\label{twotime}
g^<_{ab}&=&{\cal G}^<_{ab}+{\cal
G}^R_{ab}\Big[V_{ab}+\Sigma^R_{ab}\Big]g_{ab}^<
+{\cal G}^R_{ab}\sigma^<_{ab}G_{ab}^A
+{\cal G}^<_{ab}\Big[V_{ab}+\Sigma^A_{ab}\Big]G_{ab}^A\\[1ex]
G^A_{ab}&=&{\cal G}^A_{ab}+{\cal
G}^A_{ab}\Big[V_{ab}+\Sigma^A_{ab}\Big]\,G^A_{ab}
\label{twotime1}
\end{eqnarray}
with the definitions
${\cal G}^<_{ab}(t,t')=g^<_{a,0}(t,t')\,g^<_{b,0}(t,t')\,$,
${\cal G}^R_{ab}(t,t')=ig^R_{a,0}(t,t')\,g^R_{b,0}(t,t')\,$, and
${\cal G}^A_{ab}(t,t')=(-i)g^A_{a,0}(t,t')\,g^A_{b,0}(t,t')\,$.	All
quantities in the above	equations depend on	two	times only.	The
integration	of intermediate	times runs in the interval $[t_0,\infty]$
like in	Eqs.~(\ref{dyson1},\ref{dyson1R}).

The	zeroth,	first and
second orders for $g_{ab}^<$ with respect to the two-particle
self-energy	are	given by
\begin{eqnarray}\label{struct0}
g^{<\,(0)}_{ab}&=&{\cal	G}^<_{ab}
\,,
\\
\label{struct1}
g^{<\,(1)}_{ab}&=&{\cal	G}^R_{ab}\Big[V_{ab}
+\Sigma^R_{ab}\Big]{\cal G}_{ab}^<
+{\cal G}^R_{ab}\sigma^<_{ab}{\cal G}_{ab}^A
+{\cal G}^<_{ab}\Big[V_{ab}+\Sigma^A_{ab}\Big]{\cal	G}_{ab}^A
\,,
\\[1ex]
\label{struct2}
g^{<\,(2)}_{ab}&=&{\cal	G}^R_{ab}\Big[V_{ab}+\Sigma^R_{ab}\Big]
{\cal G}^R_{ab}\Big[V_{ab}+\Sigma^R_{ab}\Big]
{\cal G}_{ab}^<
+{\cal G}^R_{ab}\Big[V_{ab}+\Sigma^R_{ab}\Big]
{\cal G}^R_{ab}\sigma^<_{ab}{\cal G}_{ab}^A\\ \nonumber
&&+{\cal G}^R_{ab}\Big[V_{ab}+\Sigma^R_{ab}\Big]
{\cal G}^<_{ab}\Big[V_{ab}+\Sigma^A_{ab}\Big]{\cal G}_{ab}^A
+{\cal G}^R_{ab}\sigma^<_{ab}{\cal G}_{ab}^A
\Big[V_{ab}+\Sigma^A_{ab}\Big]{\cal	G}_{ab}^A\\	\nonumber
&&+{\cal G}^<_{ab}\Big[V_{ab}+\Sigma^A_{ab}\Big]{\cal G}_{ab}^A
\Big[V_{ab}+\Sigma^A_{ab}\Big]{\cal	G}_{ab}^A
\,.
\end{eqnarray}
The	self-energy	functions $\sigma_{ab}^<$ and $\Sigma_{ab}^R$,
respectively, are identified by	comparison with	the	expansion terms	of
the	ladder equation, Fig.~\ref{diagram}, then.

All	functions in
the	above equations	(\ref{twotime})	and	(\ref{twotime1}) are
understood to depend on	two	times. The key idea	in order to	achieve
such a two-time	structure of the equations is to use the semi-group
properties of the ideal	single-particle	propagators	$g_{a,0}^R$	and
$g_{a,0}^A$	(the time-local	Hartree-Fock self-energy could also	be
included). In particular one has for any time ${\bar t}$ with
$t>{\bar t}>t'$	the	following relation
\begin{equation}
g_{a,0}^R({\bf r}_1	t,{\bf r}_1't')=i\int d^3r_2\,
g_{a,0}^R({\bf r}_1t,{\bf r}_2{\bar	t})
\,g_{a,0}^R({\bf r}_2{\bar t},{\bf r}_1't')\,.
\end{equation}
There is no	time integration in	the	above equation.
Analogeously, for the advanced function	with $t<{\bar t}<t'$ one has
(integration with respect to ${\bf r}_2$ suppressed)
\begin{equation}
g_{a,0}^A(t,t')=(-i)\,g_{a,0}^A(t,{\bar	t})\,g_{a,0}^A({\bar t},t')\,.
\end{equation}
For	the	ideal one-particle correlation functions $g_{a,0}^\gtrless$,
there follows
\begin{eqnarray}
g_{a,0}^\gtrless(t,t')&=&i\,g_{a,0}^R(t,{\bar t})\,
g_{a,0}^\gtrless({\bar t},t')\,\quad\mbox{for}\quad	t>{\bar	t}\,,\\
g_{a,0}^\gtrless(t,t')&=&(-i)\,g_{a,0}^\gtrless(t,{\bar	t})\,
g_{a,0}^A({\bar	t},t')\,\quad\mbox{for}\quad {\bar t}<t'\,,\nonumber\\
g_{a,0}^\gtrless(t,t')&=&(-i^2)\,
g_{a,0}^R(t,{\tilde	t})\,g_{a,0}^\gtrless({\tilde t},{\bar t})\,
g_{a,0}^A({\bar	t},t')\,\quad\mbox{for}\quad t>{\tilde
t}\,\quad\mbox{and}\quad{\bar t}<t'\,.\nonumber
\end{eqnarray}

Proceeding in the manner presented in Appendix~\ref{orde} and
comparing the results with the anticipated structure
[Eqs.~(\ref{twotime}),(\ref{twotime1})], we	get	the
following expression for the retarded self-energy function
\begin{eqnarray}\label{retard-s}
\Sigma_{ab}^R(t,t')&=&i\Sigma^R_a(t,t')\,g^R_{b,0}(t,t') +
i\Sigma^R_b(t,t')\,g^R_{a,0}(t,t')\\
&&	+ i	g_{b,0}^R(t,t')\,
iV_{ab}^{S\,>}(t,t')\, g_{a,0}^R(t,t')+	i g_{b,0}^<(t,t')\,
iV_{ab}^{S\,R}(t,t')\, g_{a,0}^R(t,t')\nonumber
\\
&& + i g_{a,0}^R(t,t')\,
iV_{ab}^{S\,<}(t',t)\, g_{b,0}^R(t,t')+	 i g_{a,0}^<(t,t')\,
iV_{ab}^{S\,A}(t',t)\, g_{b,0}^R(t,t')\,,\nonumber
\end{eqnarray}
where the one-particle self-energies have to be	used in	first order	of
the	dynamically	screened potential $V^S$ [cf. Eq.~(\ref{sigma-one})],
i.e.
\begin{eqnarray}
\Sigma_{a}^R(t,t')&=&\Sigma_{a}^{\rm H}(t)\delta(t-t')
+iV_{aa}^{S\,>}(t,t')\,	g_a^R(t,t')+
iV_{aa}^{S\,R}(t,t')\, g_a^<(t,t')\,.
\end{eqnarray}

\begin{figure}[h]
\centerline{\psfig{figure=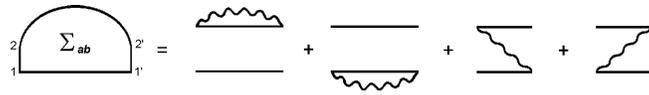,width=8.5cm,angle=0}}
\vspace*{0.5cm}
\caption[]{Diagramatic structure of	the	two-particle self-energy in
first order	in $V^S$.}\label{sigma-vs}
\end{figure}

The	correlation	function $\sigma_{ab}^<$ is	found to be
\begin{eqnarray}\label{sigma-kl}
\sigma_{ab}^<(t,t')&=&\Sigma_a^<(t,t')g^<_{b,0}(t,t')+
\Sigma_b^<(t,t')g^<_{a,0}(t,t')\\
&&+g^<_{b,0}(t,t')iV_{ab}^{S\,<}(t,t')g_{a,0}^<(t,t')
+g^<_{a,0}(t,t')iV_{ab}^{S\,>}(t',t)g_{b,0}^<(t,t')\,,\nonumber
\end{eqnarray}
where the single-particle self-energy function $\Sigma_a$ is given by
\begin{eqnarray}
\Sigma_{a}^<(t,t')&=&
iV_{aa}^{S\,<}(t,t')\, g_a^<(t,t')\,.
\end{eqnarray}
The	diagrammatic structure of the two-particle self-energy functions
is shown in	Fig.~\ref{sigma-vs}. Primarily,	these functions	consist	of
naked lines	because	we worked in first order with respect to the
dynamically	screened potential.	However, all diagrams necessary	to
dress the lines	could be found in higher orders	of the expansion, see
Appendix \ref{sec-or} and especially Fig.~\ref{classif}.

The	self-energy	functions $\Sigma^{R/A}_{ab}$ and $\sigma^<_{ab}$ are
functionals	of single-particle Green's functions. That is the reason
why	it is sufficient to	consider the two equations (\ref{twotime}) and
(\ref{twotime1}) in	order to determine the correlation function
$g_{ab}^<$.	In higher approximations, the two-particle self-energy is
expected to	be a functional	of two-particle	correlation	functions,
too. Then one would	need also the equations	for	the	other three
quantities defined in Eqs.~(\ref{g+--+}-\ref{g--++}). The full scheme
of equations reads
\begin{eqnarray}\label{fs:g-phi}
g^{\Phi}_{ab}&=&{\cal G}^{\Phi}_{ab}+{\cal
G}^R_{ab}\Big[V_{ab}+\Sigma^R_{ab}\Big]g_{ab}^{\Phi}
+{\cal G}^R_{ab}\sigma^{\Phi}_{ab}G_{ab}^A
+{\cal G}^{\Phi}_{ab}\Big[V_{ab}+\Sigma^A_{ab}\Big]G_{ab}^A\\[1ex]
&&\mbox{with}\quad \Phi=\{++,--\}; \{+-,-+\}; \{-+,+-\};\{--,++\}
\,,
\nonumber\\[1ex]
G^R_{ab}&=&{\cal G}^R_{ab}+{\cal
G}^R_{ab}\Big[V_{ab}+\Sigma^R_{ab}\Big]\,G^R_{ab}
\,,
\label{fs:gR}
\\[1ex]
G^A_{ab}&=&{\cal G}^A_{ab}+{\cal
G}^A_{ab}\Big[V_{ab}+\Sigma^A_{ab}\Big]\,G^A_{ab}
\,.			\label{fs:gA}
\end{eqnarray}
For	the	$\sigma^\phi$ see next subsection.

These equations	are	not	all	independent. $G^R_{ab}$	and	$G^A_{ab}$ are
connected by hermitean conjugation.	Further	they are linear
combinations of	the	preceding four functions $g_{ab}^{\Phi}$ according
to Eq.~(\ref{retard-G}). The system	of equations is	consistent,	i.e.,
combining the equations	for	the	$g_{ab}^\Phi$ according	to
(\ref{retard-G}) one gets the Dyson	equation (\ref{fs:gR}).

Often it is	more useful	to consider	differential equations
\begin{eqnarray}\label{kineq}
\left[i\frac{\partial}{\partial	t}-\hat
H^0_{ab}-V_{ab}\right]\,g_{ab}^\Phi(t,t')=\int d{\bar t}\,\Big[
\Sigma_{ab}^R(t,{\bar t})\,g_{ab}^\Phi({\bar t},t')	+
\sigma_{ab}^\Phi(t,{\bar t})\,G_{ab}^A({\bar t},t')\Big]\,.
\end{eqnarray}
There are additional equations for the propagator functions	$G_{ab}^R$
and	$G_{ab}^A$.	As these two functions are connected by	Hermitean
conjugation,
only the respective	equation for $G_{ab}^R$	is written down.
\begin{eqnarray}\label{diffequ}
\left[i\frac{\partial}{\partial	t}-\hat
H^0_{ab}-V_{ab}\right]\,G_{ab}^R(t,t')=\delta(t-t'\,)
+ \int d{\bar t}\,
\Sigma_{ab}^R(t,{\bar t})\,G_{ab}^R({\bar t},t')\,.
\end{eqnarray}

The	Eqs.~(\ref{fs:g-phi}--\ref{fs:gA}) and
(\ref{kineq}--\ref{diffequ}), respectively,	can	be considered as the
most important result of the present paper.	The	latter equations are
the	two-particle counterpart to	the	Kadanoff-Baym equations	in the
single-particle	case. Thus,	these equations	are	the	proper basis for
the	description	of two-particle	properties.

%%%%%%%%%%%%%%%%%%%%%%%%%%%%%%%%%%%%%%%%%%%%%%%%%%%%%%%%%%%%%%%%%%%%
\subsection{Algebraic structure	of the two-particle	self-energy}
\label{algebra-sigma}
%%%%%%%%%%%%%%%%%%%%%%%%%%%%%%%%%%%%%%%%%%%%%%%%%%%%%%%%%%%%%%%%%%%%
Similarily to the four different two-particle Green's functions
considered in Eqs.~(\ref{gsmaller},	\ref{g+--+}-\ref{g--++})
self-energy	functions $\sigma_{ab}^{\alpha,\beta,\gamma,\delta}
({\bf r}_1 {\bf	r}_2 t,	{\bf r}_1' {\bf	r}_2' t'\,)$ were introduced
in (\ref{fs:g-phi}). Within	the	present	approximation they are given
explicitely	below .
The	function $\sigma_{ab}^{++,--}$ ist just	given by
$\sigma_{ab}^<$	of Eq.~(\ref{sigma-kl}). Making	use	of
$V_{ab}^{S\,>}(1,1')=V_{ba}^{S\,<}(1',1)$, we get
\begin{eqnarray}
\label{sigma++--}
\sigma_{ab}^{++,--}
&=&	i\Big[V_{aa}^{S\,<}({\bf r}_1t,{\bf	r}_1't')
+V_{bb}^{S\,<}({\bf	r}_2t,{\bf r}_2't')
+V_{ab}^{S\,<}({\bf	r}_1t,{\bf r}_2't')
+V_{ba}^{S\,<}({\bf	r}_2t,{\bf r}_1't')\Big]
\\
&&\times g^<_{a,0}({\bf	r}_1t,{\bf r}_1't')
g_{b,0}^<({\bf r}_2t,{\bf r}_2't')
\,,
\nonumber\\
\label{sigma+--+}
\sigma_{ab}^{+-,-+}
&=&	i\Big[V_{aa}^{S\,<}({\bf r}_1t,{\bf	r}_1't')
+V_{bb}^{S\,>}({\bf	r}_2t,{\bf r}_2't')
+V_{ab}^{S\,}({\bf r}_1t,{\bf r}_2't')
+\overline{V}_{ba}^{S}({\bf	r}_2t,{\bf r}_1't')\Big]
\\
&&\times g^<_{a,0}({\bf	r}_1t,{\bf r}_1't')
g_{b,0}^>({\bf r}_2t,{\bf r}_2't')
\,,
\nonumber\\
\label{sigma-++-}
\sigma_{ab}^{-+,+-}
&=&	i\Big[V_{aa}^{S\,>}({\bf r}_1t,{\bf	r}_1't')
+V_{bb}^{S\,<}({\bf	r}_2t,{\bf r}_2't')
+\overline{V}_{ab}^{S}({\bf	r}_1t,{\bf r}_2't')
+V_{ba}^{S}({\bf r}_2t,{\bf	r}_1't')\Big]
\\
&&\times g^>_{a,0}({\bf	r}_1t,{\bf r}_1't')
g_{b,0}^<({\bf r}_2t,{\bf r}_2't')
\,,
\nonumber\\
\label{sigma--++}
\sigma_{ab}^{--,++}
&=&	i\Big[V_{aa}^{S\,>}({\bf r}_1t,{\bf	r}_1't')
+V_{bb}^{S\,>}({\bf	r}_2t,{\bf r}_2't')
+V_{ab}^{S\,>}({\bf	r}_1t,{\bf r}_2't')
+V_{ba}^{S\,>}({\bf	r}_2t,{\bf r}_1't')\Big]
\\
&&\times g^>_{a,0}({\bf	r}_1t,{\bf r}_1't')
g_{b,0}^>({\bf r}_2t,{\bf r}_2't')
\,.
\nonumber
\end{eqnarray}
The	retarded two-particle self-energy is given then	by
\begin{eqnarray}\label{retard-sigma}
\Sigma_{ab}^R({\bf r}_1	{\bf r}_2 t, {\bf r}_1'	{\bf r}_2'
t'\,)&=&\Sigma_{ab}^0({\bf r}_1	{\bf r}_2, {\bf	r}_1' {\bf r}_2'
t\,) \delta(t-t')\\
&&+\Theta(t-t')i\Big[\sigma_{ab}^{++,--}-\sigma_{ab}^{+-,-+}
-\sigma_{ab}^{-+,+-}+\sigma_{ab}^{--,++}\Big]\nonumber\,.
\end{eqnarray}
The	term which is local	in time	consists of	the	single-particle
Hartree	and	Hartree-Fock self-energies as well as the Pauli-blocking
contribution
\begin{eqnarray}\label{retard-sigma0}
\Sigma_{ab}^0({\bf r}_1	{\bf r}_2 t, {\bf r}_1'	{\bf r}_2'
t'\,)&=&\Sigma_{a}^{HF}({\bf r}_1 {\bf r}_1' t)\delta({\bf r}_2-{\bf
r}_2')+\Sigma_{b}^{HF}({\bf	r}_2 {\bf r}_2'	t)\delta({\bf r}_1-{\bf
r}_1')\\
&& +\Big[ig_a^<({\bf r}_1 t, {\bf r}_1'	t)\delta({\bf r}_2 - {\bf
r}_2')+ig_b^<({\bf r}_2	t, {\bf	r}_2' t)\delta({\bf	r}_2 - {\bf
r}_2') \Big]V_{ab}(r_1'-r_2') \nonumber
\,.
\end{eqnarray}
Inserting the expressions for the $\sigma_{ab}^\Phi$,
Eqs. (\ref{sigma++--}--\ref{sigma--++}), into Eq.~\ref{retard-sigma},
one	gets indeed	Eq.~(\ref{retard-s}).

The	advanced quantity is hermitean conjugated, and one has
\begin{eqnarray}\label{adva-sigma}
\Sigma_{ab}^A({\bf r}_1	{\bf r}_2 t, {\bf r}_1'	{\bf r}_2' t'\,)
&=&
\Sigma_{ab}^{0,\dagger}({\bf r}_1 {\bf r}_2, {\bf r}_1'	{\bf r}_2't\,)
\delta(t-t')\\
&&+\Theta(t'-t)(-i)\Big[\sigma_{ab}^{++,--}-\sigma_{ab}^{+-,-+}
-\sigma_{ab}^{-+,+-}+\sigma_{ab}^{--,++}\Big]\nonumber
\,,
\end{eqnarray}
with
\begin{eqnarray}\label{adva-sigma0}
\Sigma_{ab}^{0\dagger}({\bf	r}_1 {\bf r}_2 t, {\bf r}_1' {\bf r}_2'
t'\,)&=&\Sigma_{a}^{HF}({\bf r}_1 {\bf r}_1' t)\delta({\bf r}_2-{\bf
r}_2')+\Sigma_{b}^{HF}({\bf	r}_2 {\bf r}_2'	t)\delta({\bf r}_1-{\bf
r}_1')\\
&& +V_{ab}(r_1-r_2)
\Big[ig_a^<({\bf r}_1 t, {\bf r}_1'	t)\delta({\bf r}_2 - {\bf
r}_2')+ig_b^<({\bf r}_2	t, {\bf	r}_2' t)\delta({\bf	r}_2 - {\bf
r}_2') \Big]
\,.
\nonumber
\end{eqnarray}

Often it is	useful to consider Fourier transforms with respect to the
difference time	$t-t'$ according to
\begin{equation}
F(\omega,T)=\int d(t-t') e^{i\omega	(t-t')}	F(t-t',T)\,,
\end{equation}
with the ``macroscopic'' time $T=(t+t')/2$.	In thermodynamic
equilibrium	there is no	dependence on this time.

The	analytic properties	of the two-particle	self-energy	are	more
involved than those	of the single-particle self-energies
\cite{landau10}	because	there can occur	off-diagonal matrix	elements
already	in the spatially homogeneous case. The two-particle
self-energy	$\Sigma_{ab}^R(\omega,T)$ can be split into	hermitean and
antihermitean parts	according to
\begin{eqnarray}
\Sigma^R_{ab}(121'2',\omega	T)=\Sigma^{\cal	H}_{ab}(121'2',\omega T)+
\Sigma^{\cal A}_{ab}(121'2',\omega T)
\end{eqnarray}
with $\Sigma^{\cal H}_{ab}(121'2',\omega T)=
\Sigma^{\cal H}_{ab}(1'2'12,\omega T)^\ast$	and
$\Sigma^{\cal A}_{ab}(121'2',\omega	T)=-\Sigma^{\cal
A}_{ab}(1'2'12,\omega T)^\ast$.	These functions	can	be constructed in
the	usual way. The are connected by	Kramers-Kronig relations,
see	Appendix \ref{app-hermite}.

%%%%%%%%%%%%%%%%%%%%%%%%%%%%%%%%%%%%%%%%%%%%%%%%%%%%%%%%%%%%%%
\section{Thermodynamic equilibrium}\label{tde}
%%%%%%%%%%%%%%%%%%%%%%%%%%%%%%%%%%%%%%%%%%%%%%%%%%%%%%%%%%%%%%%
\subsection{Two-particle Dyson equation}
In thermodynamic equilibrium only the spectral properties have to be
determined,	i.e., only Eq.~(\ref{diffequ}) for the two-particle
propagator has to be considered. Its Fourier transform is
\begin{eqnarray}\label{dysoneq}
\Big[\Omega-H_{ab}^0-V_{ab}-\Sigma_{ab}^0-\Sigma_{ab}^{\rm
R\,corr}(\Omega)\Big]\,G^R_{ab}(\Omega)=1\,.
\end{eqnarray}
This equation can be called	two-particle Dyson equation	or,	likewise,
Bethe-Salpeter equation. We	want to	draw the reader's attention	on the
fact that this equation	is
given here for the function	$G_{ab}^R$,	whereas	in earlier attempts
it was tried to	formulate such an equation for the causal two-particle
GF,	\cite{zkkkr78,dukelskyschuck90},
or a function $g_{ab}^R=\Theta(t-t')(g_{ab}^>-g_{ab}^<)$,
\cite{bornath98}.

The	static part	of the two-particle	self-energy	in (\ref{dysoneq}) is
given by
%\begin{eqnarray*}
$\Sigma_{ab}^0=\Sigma_a^{\rm HF}+\Sigma_b^{\rm
HF}+(N_{ab}-1)V_{ab}$.
The	self-energy	consists of	two	different types	of terms: some do not
contain	 an	interaction	between	the	particles $a$ and $b$ whereas the
others do. The first terms are due to single-particle self-energies.

For	the	correlation	part of	the	two-particle self-energy, there	holds
the	same distinction: $\Sigma_{ab}^{R\,\rm
corr}(\Omega)$ consists	of two contributions according to
\begin{eqnarray}
\Sigma_{ab}^{R\,\rm
corr}(\Omega)=\Delta^R_{ab}(\Omega)+V_{ab}^{\rm	eff\,R}(\Omega)\,,
\end{eqnarray}
where the first	one	is due to the one-particle self-energies, whereas
the	second one describes an	effective interaction between particles
$a$	and	$b$	in the many-particle system. We	have
\begin{eqnarray}
%\Sigma_{ab}^{R,\,{\rm self}}
&&\Delta^R_{ab}({\bf p}_1 {\bf p}_2,{\bf p}_1'{\bf p}_2',\Omega)
%&=&
=
\int\limits_{-\infty}^\infty \frac{d\omega_1 d\omega_2 d\omega_3}
{(2\pi)^3}\,
\int\frac{d^3 q}{(2\pi)^3}
\frac{1}{\Omega-\omega_1-\omega_2-\omega_3+i0}
\\&&\times
\Big\{[i V_{aa}^>({\bf q},\omega_1)\, ig_a^>({\bf p}_1+{\bf	q},\omega_2)
-i V_{aa}^<({\bf q},\omega_1)\,	ig_a^<({\bf	p}_1+{\bf q},\omega_2)]
%\nonumber\\&&\quad	\times
[ig_b^>({\bf p}_2,\omega_3)-ig_b^<({\bf
p}_2,\omega_3)]\nonumber\\
&&+[i V_{bb}^>({\bf	q},\omega_1)\, ig_b^>({\bf p}_2-{\bf q},\omega_2)
-i V_{bb}^<({\bf q},\omega_1)\,	ig_b^<({\bf	p}_2-{\bf q},\omega_2)]
%\nonumber\\&&\quad	\times
[ig_a^>({\bf p}_1,\omega_3)-ig_a^<({\bf
p}_1,\omega_3)]
\Big\}
\nonumber
\\
&&\times
(2\pi)^6\, \delta({\bf p}_1-{\bf p}_1')\,
\delta({\bf	p}_2-{\bf p}_2')\nonumber \,,
\end{eqnarray}
and
\begin{eqnarray}
&&
%\Sigma_{ab}^{R,\,{\rm int}}
V_{ab}^{\rm	eff\,R}
({\bf p}_1 {\bf	p}_2,{\bf p}_1'{\bf
p}_2',\Omega)
\\
&&=(2\pi)^3\, \delta({\bf p}_1
+{\bf p}_2-{\bf	p}_1'-{\bf p}_2')
\int\limits_{-\infty}^\infty \frac{d\omega_1 d\omega_2 d\omega_3}
{(2\pi)^3}\,
\frac{1}{\Omega-\omega_1-\omega_2-\omega_3+i0}\nonumber\\
&&\times\bigg\{\Big\{
[ig_a^>({\bf p}_1,\omega_1)	- ig_a^<({\bf p}_1,\omega_1)]\,
iV_{ab}^<({\bf p}_1-{\bf p}_1',-\omega_2)\,\nonumber\\
&&\qquad+ig_a^<({\bf p}_1,\omega_1)]\,
[iV_{ab}^<({\bf	p}_1-{\bf p}_1',-\omega_2)
-iV_{ab}^>({\bf	p}_1-{\bf p}_1',-\omega_2)]\Big\}\,
[ig_b^>({\bf p}_2',\omega_3) - ig_b^<({\bf p}_2',\omega_3)]\nonumber\\
&&+\Big\{
[ig_b^>({\bf p}_2,\omega_1)	- ig_b^<({\bf p}_2,\omega_1)]\,
iV_{ab}^>({\bf p}_1-{\bf p}_1',\omega_2)\,\nonumber\\
&&\qquad+ig_b^<({\bf p}_2,\omega_1)]\,
[iV_{ab}^>({\bf	p}_1-{\bf p}_1',\omega_2)
-iV_{ab}^<({\bf	p}_1-{\bf p}_1',\omega_2)]\Big\}\,
[ig_a^>({\bf p}_1',\omega_3) - ig_a^<({\bf
p}_1',\omega_3)]\bigg\}\nonumber\,.
\end{eqnarray}
This can be	further	evaluated using	a quasiparticle	approximation
\begin{eqnarray}
\pm	i\,g_{a}^<({\bf	p},\omega)
&=&2\pi\delta(\omega-\epsilon_a(p))\,f_a(\omega)\\
i\,g_{a}^>({\bf	p},\omega)
&=&2\pi\delta(\omega-\epsilon_a(p))\,[1\pm f_a(\omega)]\,, \nonumber
\end{eqnarray}
with $f_a(\omega)=\{\exp[\beta(\omega-\mu_a]\mp	1]^{-1}$.
The	correlation	functions $V_{ab}^\gtrless$
can	be expressed in	terms of the dielectric	function
$\varepsilon(\omega)$ and Bose functions
$n_B(\omega)=[\exp(\beta\omega)-1]^{-1}$ according to
\cite{stolz,kker86}
\begin{eqnarray}
i\,V_{ab}^<({\bf q},\omega)
&=&-2 V_{ab}(q)	{\rm Im}\,{\varepsilon^{R}}^{-1}
({\bf q},\omega)\,n_B(\omega)
\,,
\\
i\,V_{ab}^>({\bf q},\omega)
&=&-2 V_{ab}(q)	{\rm Im}\,{\varepsilon^{R}}^{-1}
({\bf q},\omega)\,[1+n_B(\omega)]
 \nonumber\,.
\end{eqnarray}
For	the	function $\Delta_{ab}^R(\Omega)$ there follows
\begin{eqnarray}\label{delta}
&&\Delta^R_{ab}({\bf p}_1 {\bf p}_2,{\bf p}_1'{\bf p}_2',\Omega)
=(2\pi)^6\,
%\delta({\bf p}_1-{\bf p}_1')\,
%\delta({\bf p}_2-{\bf p}_2')
\delta_{{\bf p}_1,{\bf p}_1'}
\delta_{{\bf p}_2,{\bf p}_2'}
\int\frac{d^3 q}{(2\pi)^3}
\int\limits_{-\infty}^\infty \frac{d\omega_1}
{\pi}\,
\Big[-{\rm Im}\,
%\varepsilon^{R\,-1}({\bf q},\omega_1)
{\varepsilon^{R}}^{-1}({\bf	q},\omega_1)
\Big]\\
&&\qquad\times\bigg\{
V_{aa}({\bf	q})\frac{1\pm f_a({\bf p}_1+{\bf q})+n_B(\omega_1)}
{\Omega-\omega_1-\epsilon_a({\bf p}_1+{\bf q})-\epsilon_b({\bf
p}_2)+\!i0}
+(a\longleftrightarrow b\,,\,1\longleftrightarrow2\,)
\bigg\}
\nonumber\\[1ex]
&&=(2\pi)^6\,\delta_{{\bf p}_1,{\bf	p}_1'}
\delta_{{\bf p}_2,{\bf p}_2'}
\Delta^R_{ab}({\bf p}_1'{\bf p}_2',\Omega)\nonumber
 \,.
\end{eqnarray}
In the last	line of	this equation, the function
$\Delta^R_{ab}({\bf	p}_1'{\bf p}_2',\Omega)$ was introduced	which is
the	sum	of the single-particle self-energies (in $V^S$ approximation)
to be taken	off-shell
\begin{equation}
\Delta^R_{ab}({\bf p}_1'{\bf p}_2',\Omega)=
\Sigma^R_a\Big({\bf	p}_1',\Omega-\epsilon_b({\bf p}_2')\Big)+
\Sigma^R_b\Big({\bf	p}_2',\Omega-\epsilon_a({\bf p}_1')\Big)
\,
.
\end{equation}

The	other contribution to the two-particle self-energy,
$V_{ab}^{\rm eff}$,	is given by
\begin{eqnarray}\label{veff}
&&V_{ab}^{\rm eff\,R}
({\bf p}_1 {\bf	p}_2,{\bf p}_1'{\bf	p}_2',\Omega)
= (2\pi)^3\,
%\delta({\bf p}_1+{\bf p}_2-{\bf p}_1'-{\bf	p}_2')
\delta_{{\bf p}_1+{\bf p}_2,{\bf p}_1'+{\bf	p}_2'}
\int\limits_{-\infty}^\infty \frac{d\omega_1}{\pi}\,
\Big[
-{\rm Im}\,{\varepsilon^{R}}^{-1}
({\bf p}_1-{\bf	p}_1',\omega_1)
\Big]
\nonumber\\
&&\qquad\times\bigg\{
V_{ab}({\bf	p}_1-{\bf p}_1')\frac{1\pm f_a({\bf	p}_1)+n_B(\omega_1)}
{\Omega-\omega_1-\epsilon_a({\bf p}_1)-\epsilon_b({\bf p}_2')+\!i0}
\,+\,(a\longleftrightarrow b\,,\,1\longleftrightarrow2\,)
\bigg\}
\,.
\end{eqnarray}
These two contributions	to the two-particle	self-energy	look very
similar. Replacing the Coulomb potential $V_{ab}$ by $z_az_bV$ with
$z_a,z_b$ being	the	charge numbers,	one	can	see	that for particles
attracting each	other, there is	a compensation between these two
functions. This	is especially to be	seen considering the functions
integrated with	respect	to ${\bf p}_1$ and ${\bf p}_2$
\begin{eqnarray}
&&\int\frac{d^3p_1}{(2\pi)^3}\int\frac{d^3p_2}{(2\pi)^3}
\Big[
\Delta^R_{ab}({\bf p}_1	{\bf p}_2,{\bf p}_1'{\bf p}_2',\Omega)
+V_{ab}^{\rm eff\,R}({\bf p}_1 {\bf	p}_2,{\bf p}_1'{\bf
p}_2',\Omega)
\Big]\nonumber\\
&=&\Big(z_a+z_b\Big)
\int\frac{d^3 q}{(2\pi)^3}
\int\limits_{-\infty}^\infty \frac{d\omega_1}
{\pi} \,
\Big[-{\rm Im}\,
%\varepsilon^{R\,-1}({\bf q},\omega_1)
{\varepsilon^{R}}^{-1}({\bf	q},\omega_1)
\Big]\\
&&\qquad\times\bigg\{
z_a\,V({\bf	q})\frac{1\pm f_a({\bf p}_1'+{\bf q})+n_B(\omega_1)}
{\Omega-\omega_1-\epsilon_a({\bf p}_1'+{\bf	q})-\epsilon_b({\bf
p}_2')+\!i0}
+(a\longleftrightarrow b\,,\,1\longleftrightarrow2\,)
\bigg\}
\,.		   \nonumber
\end{eqnarray}
In the case	of a symmetrical plasma, $z_a=-z_b$, the right hand	side
of the above equation vanishes and it follows
\begin{eqnarray}
\Delta^R_{ab}({\bf p}_1'{\bf p}_2',\Omega)
=-\int\frac{d^3p_1}{(2\pi)^3}\int\frac{d^3p_2}{(2\pi)^3}\,
V_{ab}^{\rm	eff\,R}({\bf p}_1 {\bf p}_2,{\bf p}_1'{\bf
p}_2',\Omega)\,.
\end{eqnarray}

These expression for the two-particle self-energy have to be compared
with
the	results	of former papers \cite{zkkkr78,kkr83}. The notations are
slightly different in comparison with ours,	so one should compare the
expressions	of the effective hamiltonians. The total hamiltonian is
$H_{ab}^0+V_{ab}+H_{ab}^{\rm pl}(\omega)$ with $H_{ab}^0+V_{ab}$ being
the	hamiltonian	of the isolated	pair of
particles whereas the medium-dependent part	of the Hamiltonian is
denoted	by $H_{ab}^{\rm	pl}(\omega)$. In the present paper this	latter
quantity is	given by
\begin{eqnarray}
H_{ab}^{\rm	pl}(\omega)
&=&\Sigma_{ab}^0
+\Sigma_{ab}^{\rm corr}(\omega)\\
&=&\Sigma_a^{\rm HF}+\Sigma_b^{\rm HF}+N_{ab}V_{ab}-V_{ab}
+\Delta^R_{ab}(\Omega)+V_{ab}^{\rm eff\,R}(\Omega)
\,,
\nonumber
\end{eqnarray}
where $\omega$ is to be	understood as an parameter.

The	differences	consist	in the following: (i) now there	a no
additional static parts	beyond the Hartree-Fock	level,
(ii) no	division by	Pauli-blocking terms occurs. Both things seem to
be produced	artificially by	adopting a closed equation for the wrong
quantity.

On the other hand, the same	results	as before are received in the
nondegenerate case.

%%%%%%%%%%%%%%%%%%%%%%%%%%%%%%%%%%%%%%%%%%%%%%%%%%%%%%%%%%%%%%%%%%%%%%
\subsection{Limiting cases}
%%%%%%%%%%%%%%%%%%%%%%%%%%%%%%%%%%%%%%%%%%%%%%%%%%%%%%%%%%%%%%%%%%%%%%
It is interesting to study some	limiting cases of our expressions.
First, in the nondegenerate	case the one-particle distribution
functions in Eqs.~(\ref{delta})	and	(\ref{veff}) can be	neglected.
The	result is in aggreement	with the nondegenerate limit of	the	former
approaches using the Shindo	approximation
\cite{zkkkr78,kkr83,kker86,z88}.

The	second important limiting case is that of statical screening.
Following Zimmermann \cite{z88}	we consider	the	case that the
excitation energy into a pair of two free particles,
$\epsilon_a({\bf p}_1)+\epsilon_b({\bf p}_2')-\Omega$, is small	in
comparison with	the	energy $\omega_1$ occuring in the dielectric
function. This could be	a reasonable approximation for excited states.
Then we	have
\begin{eqnarray}\label{delta-stat}
&&\Delta^R_{ab}({\bf p}_1 {\bf p}_2,{\bf p}_1'{\bf p}_2',\Omega)
=(2\pi)^6\,	\delta({\bf	p}_1-{\bf p}_1')\,
\delta({\bf	p}_2-{\bf p}_2')
\int\frac{d^3 q}{(2\pi)^3}
\int_{-\infty}^\infty \frac{d\omega_1}
{\pi}\,
\Big[-{\rm Im}\,
%\varepsilon^{R\,-1}({\bf q},\omega_1)
{\varepsilon^{R}}^{-1}({\bf	q},\omega_1)
\Big]\nonumber\\
&&\qquad\times\bigg\{
V_{aa}({\bf	q})\frac{1\pm f_a({\bf p}_1+{\bf q})+n_B(\omega_1)}
{-\omega_1+\!i0}
+(a\longleftrightarrow b\,,\,1\longleftrightarrow2\,)
\bigg\}
\\
&&=(2\pi)^6\, \delta({\bf p}_1-{\bf	p}_1')\,
\delta({\bf	p}_2-{\bf p}_2')
\int\frac{d^3 q}{(2\pi)^3}
\Bigg\{
\Big[V^S_{aa}({\bf q},\Omega=0)-V_{aa}({\bf	q})\Big]\Big[\pm f_a({\bf
p}_1+{\bf q}) +	\frac{1}{2}\Big]\nonumber\\
&&+\Big[V^S_{bb}({\bf q},\Omega=0)-V_{bb}({\bf q})\Big]\Big[\pm	f_b({\bf
p}_2+{\bf q}) +	\frac{1}{2}\Big]
\Bigg\}
 \,,\nonumber
\end{eqnarray}
and	for	the	effective interaction term
\begin{eqnarray}\label{veff-stat}
&&V_{ab}^{\rm eff\,R}({\bf p}_1	{\bf p}_2,{\bf p}_1'{\bf
p}_2',\Omega)
= (2\pi)^3\, \delta({\bf p}_1
+{\bf p}_2-{\bf	p}_1'-{\bf p}_2')
\int\nolimits_{-\infty}^\infty \frac{d\omega_1}{\pi}\,
\Big[-{\rm Im}\,
%\varepsilon^{R\, -1}
{\varepsilon^{R}}^{-1}
({\bf p}_1-{\bf
p}_1',\omega_1)\Big]\nonumber\\
&&\qquad\bigg\{
V_{ab}({\bf	p}_1-{\bf p}_1')\frac{1\pm f_a({\bf	p}_1)+n_B(\omega_1)}
{-\omega_1+\!i0}
\,+\,(a\longleftrightarrow b\,,\,1\longleftrightarrow2\,)
\bigg\}\\
&&=
(2\pi)^3\, \delta({\bf p}_1
+{\bf p}_2-{\bf	p}_1'-{\bf p}_2')
\Big[V^S_{ab}({\bf p}_1-{\bf p}_1',\Omega=0)-V_{ab}({\bf p}_1-{\bf p}_1')
\Big]
\Big[1 \pm f_a({\bf	p}_1)\pm f_b({\bf p}_2)	\Big]
\,.\nonumber
\end{eqnarray}
Here it	was	used that ${\rm	Im}	\varepsilon^{-1}$ is an	odd	function
and	that the even part of the Bose function	$n_B(\omega)$ is $-1/2$.
Further	one	has
\begin{equation}
V^{S\,R}_{ab}({\bf q},\Omega)=V_{ab}({\bf q})\Bigg\{1-\int
\frac{d\omega_1}{\pi}\frac{{\rm	Im}
\varepsilon^{-1}({\bf q},\omega_1)}{\Omega-\omega_1+i0}
\Bigg\}\,.
\end{equation}

The	terms in Eq. (\ref{delta-stat})	containing
single-particle	distribution functions and Eq. (\ref{veff-stat})
can	be combined	with
$\Sigma_{ab}^{HF}$ and $N_{ab}\,V_{ab}$, respectively, to give
functionals	$\Sigma_{ab}^{HF}\Big\{	V^S(\Omega =0 \Big\}$ and
$N_{ab}\,V^S_{ab}(\Omega=0)$ of	the	screened potentials	in the static
limit. The remaining terms in Eq. (\ref{delta-stat}) give a	constant
term.

The	Dyson equation,	Eq.~(\ref{dysoneq}), can be	written	then
in the following form
\begin{eqnarray}
\Big[\Omega-H_{ab}^{\rm	eff}\Big]\,G^R_{ab}(\Omega)=1\,,
\end{eqnarray}
with the effective plasma Hamilton operator
\begin{eqnarray}
H_{ab}^{\rm	eff}&=&H_{ab}^0
+
\sum\limits_{c=a,b}\Sigma_c^{\rm HF}\bigg\{V^S(\Omega=0)\bigg\}
+N_{ab}\,V_{ab}^S(\Omega=0)
\\
\nonumber
&+&\frac{1}{2}\sum\limits_{c=a,b}
\int \frac{d^3
q}{(2\pi)^3}
\Big[V_{cc}^S(q,0)-V_{cc}(q)\Big]
\,.
\end{eqnarray}
Considering	this effective Hamiltonian for a nondegenerate system, one
can	write (with	$V_{ab}\rightarrow z_a z_b V$)
\begin{eqnarray}\label{nichtent}
H_{ab}^{\rm	eff}&=&H_{ab}^0
+z_a z_b V(r) +	z_az_b\,\Big[V^S(r,\Omega=0)-V(r)\Big]
\\
&+&\frac{1}{2}\Big[z_a^2+z_b^2\Big]\,\Big[V^S(0,\Omega=0)-V(0)\Big]
\,.\nonumber
\end{eqnarray}
For	a symmetrical plasma with $z_a=-z_b=1$ this	leads to
\begin{eqnarray}
H_{ab}^{\rm	eff}=H_{ab}^0
- V(r) - \bigg\{\Big[V^S(r,\Omega=0)-V(r)\Big]
-\,\Big[V^S(0,\Omega=0)-V(0)\Big]\bigg\}\,,
\end{eqnarray}
i.e., at small distances the last two terms	compensate each	other to a
large extent and the interaction is	given by the Coulomb potential
\begin{eqnarray}
\lim_{r\rightarrow 0} H_{ab}^{\rm eff}=H_{ab}^0
- V(r) + O(r^2)
\,.
\end{eqnarray}

Adopting for $V^S(r,\Omega=0)$ the statically screened Debye
potential $V^D(r)$,
\begin{eqnarray}
V^D(r)=\frac{e^2}{r}\exp{(-\kappa r)}\,;\quad V^D(q)=
\frac{4\pi e^2}{q^2+\kappa^2}\,,
\end{eqnarray}
one	gets for the Hamiltonian in	(\ref{nichtent})
\begin{eqnarray}
H_{ab}^{\rm	eff}=H_{ab}^0
+z_a z_b V^D(r)
-\frac{1}{2}\Big(z_a^2+z_b^2\Big)\,\kappa e^2
\,.
%\nonumber\\
\end{eqnarray}
The	two	last terms combined	give the well-known	effective potential	of
Ecker-Weizel type \cite{ew56} which	has	been used frequently in	order to
determine energies and wave	functions of bound states in a plasma
environment	\cite{kkr83,kker86,sb93,bs93}.

%%%%%%%%%%%%%%%%%%%%%%%%%%%%%%%%%%%%%%%%%%%%%%%%%%%%%%%%%%%%%%%%%%%
\subsection{Effective wave equation	and	two-particle energies}
%%%%%%%%%%%%%%%%%%%%%%%%%%%%%%%%%%%%%%%%%%%%%%%%%%%%%%%%%%%%%%%%%%%
The	Eq.~(\ref{dysoneq})	which determines the two-particle propagator
$G_{ab}^R$ was written down	in an operator form. Using a representation
one	gets a matrix equation.	In order to	solve this equation	it is
favorable to use a representation in which diagonal	elements are the
main contribution. We follow here Kilimann et al. \cite{kkr83},
however, now it	is not possible	to achieve symmetric real and imaginary
parts of the effective hamiltonian simply by multiplying with factors
$N_{ab}^{\pm 1/2}$.	Therefore the hamiltonian is split into	hermitean
and	antihermitean parts, see Appendix \ref{app-hermite}. This leads	to
\begin{eqnarray}\label{op}
\big[\Omega-H_{ab}^{\cal H}(\Omega)	\big]G_{ab}^{R}(\Omega)-H_{ab}^{\cal
A}(\Omega)G_{ab}^R(\Omega)=1\,,
\end{eqnarray}
with $H_{ab}^{\cal H}(\Omega)=H_a+H_b+V_{ab}+\Sigma_{ab}^{0\,{\cal
H}}+\Sigma_{ab}^{\rm corr\,{\cal H}}(\Omega)$.
The	eigenvalue problem of the hermitean	part of	the	hamiltonian	reads
\begin{eqnarray}
H_{ab}^{\cal H}(\Omega)\,\Big|nP,\Omega\Big\rangle
=E_{nP}(\Omega)\,\Big|nP,\Omega\Big\rangle\,,
\end{eqnarray}
The	eigenstates	$\Big|nP,\Omega\Big\rangle$	where $nP$ denote the
quantum	numbers	and	$\omega$ is	a real parameter can be	used as	a
orthonormal	basis. The eigenvalues $E_{nP}(\Omega)$	of this	effective
Schr"odinger equation are not yet the spectrum of two-particle
excitions \cite{kkr83}.	The	latter follows from	the	spectral function
$A_{ab}$.

In the representation with respect to the eigenstates
$\Big|nP,\Omega\Big\rangle$, Eq.~\ref{op} reads	(conservation of
center-of-mass momentum	already	taken into account)
\begin{eqnarray}\label{op1}
\big[\Omega-E_{nP}(\Omega)\big]G_{nn'}^R(P,\Omega)-\sum_m
H_{nm}^{\cal A}(P,\Omega)G_{mn'}^R(P,\Omega)=\delta_{nn'}\,.
\end{eqnarray}
In the following it	is assumed that	nondiagonal	matrix
elements of	the	antihermitean part of the effective	hamiltonian
are	small. Then	the	Eq.~\ref{op1} has the approximate solution
\begin{eqnarray}
G_{nn'}^{R}(P,\Omega)&=&\frac{\delta_{nn'}}
{\Omega+i0-E_{nP}(\Omega)+i\Gamma_{nn}(P\Omega)}
\\
&&+\frac{-i\Gamma_{nn'}(P\Omega)(1-\delta_{nn'})}
{[\Omega+i0-E_{nP}(\Omega)+i\Gamma_{nn}(P\Omega)]
[\Omega+i0-E_{n'P}(\Omega)+i\Gamma_{n'n'}(P\Omega)]}\nonumber
\,,
\end{eqnarray}
where it was introduced	$\Gamma_{nn'}(P,\Omega)=iH_{nn'}^{\cal
A}(P,\Omega)$.
For	the	coherent part of the spectral function there follows
\begin{eqnarray}
A_{nn}(P,\Omega)&=&\frac{2\Gamma_{nn}(P\Omega)}
{[\Omega-E_{nP}(\Omega)]^2+\Gamma^2_{nn}(P\Omega)}\,.
\end{eqnarray}
According to this equation,	the	spectrum of	the	two-particle
excitations	is given by	the	roots ${\tilde E}_{nP}$	of
\begin{eqnarray}
\Omega=E_{nP}(\Omega)
\end{eqnarray}
whereas	the	damping	is given by	$\Gamma_{nn}(P,{\tilde E}_{nP})$
\cite{kkr83}.
%%%%%%%%%%%%%%%%%%%%%%%%%%%%%%%%%%%%%%%%%%%%%%%%%%%%%%%%%%%%%%%%%%%
\section{Summary and Conclusion}
%%%%%%%%%%%%%%%%%%%%%%%%%%%%%%%%%%%%%%%%%%%%%%%%%%%%%%%%%%%%%%%%%%%
Starting from the nonequilibrium Bethe-Salpeter	equation in	the
dynamically	screened approximation,	we have	derived	a set of
nonequilibrium Dyson equations for two-time	two-particle
correlation	functions. The two-time	structure of these equations was
achieved in	an exact way using the semi-group properties of	the
ideal one-particle Green's functions. The prize	one	has	to pay for
this simpler structure of the equation is that the two-particle
self-energy	in the Dyson equation consists now of irreducible diagrams
in all orders with respect to the dynamically screened potential (in
some sence this	is similar to the transition from Feynman diagrams to
Goldstone diagrams \cite{blaizot86}). Irreducibility means here	that a
diagram	cannot be cut with respect to a	pair of	single-particle	lines
which begin	at equal times and end at equal	times, i.e., two ore more
interaction	potentials have	some overlap in	time.

For	the	further	considerations we have restricted ourselves	to a
two-particle self-energy in	first order	with respect to	the	screened
potential.
The	algebraic structure	of the equations is	not	affacted by
this approximation.	It was shown that there	is a set of	equations for
four two-time correlation functions. This generalizes the couple of
Kadanoff-Baym equations	for	the	one-particle correlation functions
$g^\gtrless$ ($g_{+-}$ and $g_{-+}$, respectively).	In analogy to the
single-particle	case there is no closed	equation for the correlation
functions but always a coupling	to other correlation functions.	Only
for	two	certain	functions $G_{ab}^{R/A}$ there exist closed	equations.
Thus, these	functions are the two-particle generalization of the
retarded (advanced)	commutator Green's functions $g_a^{R/A}$ in	the
single-particle	case and just these	functions describe the propagation
of a pair of particles in the nonequilibrium many-particle system.

The	case of	thermodynamic equilibrium was considered in	some detail	in
order to show the differences to former	approaches.	In former attempts
\cite{kkk77,zkkkr78} there were	anticipated	closed equations for the
causal two-time	two-particle Green's function. These equations were
enforced by	the	Shindo approximation.
The	expressions	for	the	effective hamiltonian are the same as those	of
the	present	paper only for the case	of a nondegenerate system. The
agreement in this special case is easy to understand taking	into
account	that the difference	between	the	used functions is of higher
order in the density.

For	arbitrary degeneracy there are clear differences between the
former results and ours. In	the	present	results	there is no	division
by Pauli-blocking terms. The only intrinsic	static contributions of
the	effective Hamiltonian (the two-particle	self-energy) are the
Hartree-Fock single-particle self-energies and the Pauli-blocked
basic potential.

We can conclude	that the proper	generalization of the Kadanoff-Baym
equations for two-particle functions is	given by the system	of
equations (\ref{kineq})	and	(\ref{diffequ}). The algebraic structure
of these equations was identified starting from	a concrete
approximation, the dynamically screened	ladder equation. More general
considerations how the self-energy functions can be	determined in
higher approximations will be presented	in a subsequent	paper
\cite{bor99}.

%%%%%%%%%%%%%%%%%%%%%%%%%%%%%%%%%%%%%%%%%%%%%%%%%%%%%%%%%%%%%%%%%%%
\acknowledgements
%%%%%%%%%%%%%%%%%%%%%%%%%%%%%%%%%%%%%%%%%%%%%%%%%%%%%%%%%%%%%%%%%%%
We are grateful	to W.-D. Kraeft	and	G. R\"opke for helpful
discussons.

This work has been done	under the auspices of the
Sonderforschungsbereich	``Kinetics of partially	ionized	plasmas''.

%%%%%%%%%%%%%%%%%%%%%%%%%%%%%%%%%%%%%%%%%%%%%%%%%%%%%%%%%%%%%%%%%%%
\appendix
%%%%%%%%%%%%%%%%%%%%%%%%%%%%%%%%%%%%%%%%%%%%%%%%%%%%%%%%%%%%%%%%%%%
\section{Evaluation	of dynamically screened	ladder terms}\label{orde}
%%%%%%%%%%%%%%%%%%%%%%%%%%%%%%%%%%%%%%%%%%%%%%%%%%%%%%%%%%%%%%%%%%%
The	aim	of this	appendix is	to show	the	evaluation of the
lowest-order terms in the dynamically screened ladder equation.
Single-particle	self-energy	contributions and interaction terms	have
to be treated on equal footing.	Special	attention is paid to the
transformation into	a structure	involving two-particle quantities
which depend on	two	times only.	The	analysis is	made here for the
expansion of the function $g^<_{ab}=g_{ab}^{++--}$.	Similar
considerations are possible	for	the	other three	functions
$g_{ab}^{+--+}$, $g_{ab}^{-++-}$, and $g_{ab}^{--++}$. This	is
scetched in	subsection \ref{andere}.
%%%%%%%%%%%%%%%%%%%%%%%%%%%%%%%%%%%%%%%%%%%%%%%%%%%%%%%%%%%%%%%%%%%%%%
\subsection{First-order	contributions}
%%%%%%%%%%%%%%%%%%%%%%%%%%%%%%%%%%%%%%%%%%%%%%%%%%%%%%%%%%%%%%%%%%%%%%
There are three	diagrams of	first order	with respect to	the
dynamically	screened interaction $V_{ab}^S$, see Fig.~\ref{diagram}.
Two	terms have single-particle self-energy insertions of particles $a$
and	$b$, respectively. The third one is	a ladder diagram with one
rung.

The	first term with	a self-energy insertion	for	particle $a$ is	given
simply by [cf. Eq.~\ref{dyson1}]
\begin{eqnarray}\label{I1}
I_1^{(1)}(t,t')&=&\int\limits_{t_0}^\infty dt_1	d{\bar t}_1\,
\Big[g^<_{a,0}(t,t_1)\,
\Sigma_a^A(t_1,{\bar t}_1)\,g^A_{a,0}({\bar	t}_1,t')
+ g^R_{a,0}(t,t_1)\,\Sigma_a^<(t_1,{\bar t}_1)\,
g^A_{a,0}({\bar	t}_1,t')\\
&&\qquad\quad +	g^R_{a,0}(t,t_1)\,\Sigma_a^R(t_1,{\bar t}_1)\,
g^<_{a,0}({\bar	t}_1,t')\Big]\,g^<_{b,0}(t,t')\nonumber\,.
\end{eqnarray}
In order to	achieve	the	anticipated	structure, one can use the
semi-group properties for the correlation function $g^<_{b,0}(t,t')$.
In the first term on the right hand	side, for instance,	there holds
$t_1<{\bar t}_1<t'$, enforced by the advanced functions
$\Sigma_a^A(t_1,{\bar t}_1)\,g^A_{a,0}({\bar t}_1,t')$.
For	this case we can use in	Eq.~(\ref{I1})
$g^<_{b,0}(t,t')=g^<_{b,0}(t,t_1)\,(-i)g^A_{b,0}(t_1,{\bar t}_1)
\,(-i)g^A_{b,0}({\bar t}_1,t')$. Treating the other	two	terms in a
similar	way	one	gets
\begin{eqnarray}\label{I1zz}
I_1^{(1)}(t,t')&=&\int\limits_{t_0}^\infty dt_1	d{\bar t}_1\,
\Big\{g^<_{a,0}(t,t_1)\,g^<_{b,0}(t,t_1)\,
\Big[(-i)\Sigma_a^A(t_1,{\bar t}_1)\,g^A_{b,0}(t_1,{\bar t}_1)\Big]\,
(-i)g^A_{a,0}({\bar	t}_1,t')\,g^A_{b,0}({\bar t}_1,t')
\nonumber
\\
&&\qquad + ig^R_{a,0}(t,t_1)\,g^R_{b,0}(t,t_1)\,
\Big[\Sigma_a^<(t_1,{\bar t}_1)\,g^<_{b,0}(t_1,{\bar t}_1)\Big]\,
(-i)g^A_{a,0}({\bar	t}_1,t')\,g^A_{b,0}({\bar t}_1,t')
\\[1ex]
&&\qquad + ig^R_{a,0}(t,t_1)\,g^R_{b,0}(t,t_1)\,
\Big[i\Sigma_a^R(t_1,{\bar t}_1)\,g^R_{b,0}(t_1,{\bar t}_1)\Big]\,
g^<_{a,0}({\bar	t}_1,t')\,g^<_{b,0}({\bar t}_1,t')\Big\}\,.\nonumber
\end{eqnarray}
This fits into the structure
\begin{eqnarray}
{\cal G}^<_{ab}\,\Sigma^A_{ab}\,{\cal G}_{ab}^A
+{\cal G}^R_{ab}\,\sigma^<_{ab}\,{\cal G}_{ab}^A
+{\cal G}^R_{ab}\Sigma^R_{ab}{\cal G}_{ab}^<
\,.
\end{eqnarray}
The	term $I_1^{(2)}$ containing	a self-energy insertion	for	the	other
particle of	species	$b$	has	a similar shape
\begin{eqnarray}\label{I2zz}
I_1^{(2)}(t,t')&=&\int\limits_{t_0}^\infty dt_2	d{\bar t}_2\,
\Big\{g^<_{a,0}(t,t_2)\,g^<_{b,0}(t,t_2)\,
\Big[(-i)g^A_{a,0}(t_2,{\bar t}_2)\,\Sigma_b^A(t_2,{\bar t}_2)\Big]\,
(-i)g^A_{a,0}({\bar	t}_2,t')\,g^A_{b,0}({\bar t}_2,t')
\nonumber
\\
&&\qquad + ig^R_{a,0}(t,t_2)\,g^R_{b,0}(t,t_2)\,
\Big[g^<_{a,0}(t_2,{\bar t}_2)\,\Sigma_b^<(t_2,{\bar t}_2)\Big]\,
(-i)g^A_{a,0}({\bar	t}_2,t')\,g^A_{b,0}({\bar t}_2,t')
\\[1ex]
&&\qquad + ig^R_{a,0}(t,t_2)\,g^R_{b,0}(t,t_2)\,
\Big[ig^R_{a,0}(t_2,{\bar t}_2)\,\Sigma_b^R(t_2,{\bar t}_2)\Big]\,
g^<_{a,0}({\bar	t}_2,t')\,g^<_{b,0}({\bar t}_2,t')\Big\}\,.\nonumber
\end{eqnarray}

The	third first-order term in the perturbation expansion is	the	ladder
term. Each of the two vertices can have	the	Keldysh	indices	$+$	and
$-$. Thus one gets the following four terms
\begin{eqnarray}
I_1^{(3)}(t,t')&=&i\int	d{\bar t}_1d{\bar t}_2\,
\Big[ g^{}_{a,0}(t,\bar{t}_1)\,g^{}_{b,0}(t,\bar{t}_2)\,
V_{ab}^S({\bar t}_1,{\bar t}_2)\,
g^<_{a,0}(\bar{t}_1,t')\,g^<_{b,0}(\bar{t}_2,t')
\\[1ex]
&&-g^<_{a,0}\,g^{}_{b,0}\,V_{ab}^{S\,>}\,{\bar g}^{}_{a,0}\,g^<_{b,0}
-g^{}_{a,0}\,g^<_{b,0}\,V_{ab}^{S\,<}\,g^<_{a,0}\,{\bar	g}^{}_{b,0}
%\nonumber\\&&
+g^<_{a,0}\,g^<_{b,0}\,{\bar V}_{ab}^{S}\,
{\bar g}^{}_{a,0}\,{\bar g}^{}_{b,0}\Big] \,.	  \nonumber
\end{eqnarray}
The	causal and anticausal Green's functions	can	be eliminated in
favour of retarded and advanced	GF's, see Eq.~(\ref{retardGF}),
\begin{eqnarray}\label{I3-kleiner}
I_1^{(3)}&=&g^{R}_{a,0}\,g^{R}_{b,0}\,iV_{ab}^{S}\,g^{<}_{a,0}
\,g^<_{b,0}
+g^{R}_{a,0}\,g^{<}_{b,0}\,iV_{ab}^{S\,R}\,g^{<}_{a,0}\,g^<_{b,0}
+g^{<}_{a,0}\,g^{R}_{b,0}\,iV_{ab}^{S\,A}\,g^{<}_{a,0}\,g^<_{b,0}
\\[2ex]
&&+g^{R}_{a,0}\,g^{<}_{b,0}\,iV_{ab}^{S\,<}\,g^{<}_{a,0}\,g^{A}_{b,0}
+g^{<}_{a,0}\,g^{R}_{b,0}\,iV_{ab}^{S\,>}\,g^{A}_{a,0}\,g^{<}_{b,0}
\nonumber\\[2ex]
&&+g^{<}_{a,0}\,g^{<}_{b,0}\,iV_{ab}^{S\,R}\,g^{A}_{a,0}\,g^{<}_{b,0}
+g^{<}_{a,0}\,g^{<}_{b,0}\,iV_{ab}^{S\,A}\,g^{<}_{a,0}\,g^{A}_{b,0}
+g^{<}_{a,0}\,g^{<}_{b,0}\,i{\bar V}_{ab}^{S}\,g^{A}_{a,0}
\,g^{A}_{b,0}
\,.\nonumber
\end{eqnarray}
There are three	classes	of terms in	the	above equation:	(i)	terms
ending with	a product $g^{<}_{a,0}\,g^<_{b,0}$,	(ii) terms beginning
with one retarded function and ending with one advanced	function, and
(iii) terms	beginning with $g^{<}_{a,0}\,g^<_{b,0}$.

The	further	procedure is presented in detail for the first term	in the
above equation
\begin{eqnarray}
&&\int d{\bar t}_1d{\bar t}_2\,
g^{R}_{a,0}(t,\bar{t}_1)\,g^{R}_{b,0}(t,\bar{t}_2)\,
iV_{ab}^S({\bar	t}_1,{\bar t}_2)\,
g^<_{a,0}(\bar{t}_1,t')\,g^<_{b,0}(\bar{t}_2,t')\\[1ex]
&=&\int	d{\bar t}_1d{\bar t}_2\,
ig^{R}_{a,0}(t,\bar{t}_1)\,g^{R}_{b,0}(t,\bar{t}_2)\,
\Big[V_{ab}\delta({\bar	t}_1-{\bar t}_2)+\Theta({\bar t}_1-{\bar t}_2)
V_{ab}^{S\,>}({\bar	t}_1,{\bar t}_2)\nonumber\\
&&\qquad\quad +\Theta({\bar	t}_2-{\bar t}_1)
V_{ab}^{S\,<}({\bar	t}_1,{\bar t}_2)\Big]
\,g^<_{a,0}(\bar{t}_1,t')\,g^<_{b,0}(\bar{t}_2,t')
\,.
\nonumber
\end{eqnarray}
The	Heaviside functions	allow it to	use	the	semi-group property	in
certain	functions $g^R$	and	$g^<$, respectively, in	the	following
manner
\begin{eqnarray}
&&\int d{\bar t}_1d{\bar t}_2\,\bigg\{
\Big[ig^{R}_{a,0}(t,\bar{t}_1)\,g^{R}_{b,0}(t,\bar{t}_2)\Big]\,
\Big[ V_{ab}\delta({\bar t}_1-{\bar	t}_2)\Big]
\,\Big[g^<_{a,0}(\bar{t}_1,t')\,g^<_{b,0}(\bar{t}_2,t')\Big]\\
&&\qquad +\Big[ig^{R}_{a,0}(t,\bar{t}_1)\,g^{R}_{b,0}(t,\bar{t}_1)
\Big]\,
\Big[ig^{R}_{b,0}({\bar	t}_1,{\bar t}_2)\,
iV_{ab}^{S\,>}({\bar t}_1,{\bar	t}_2)
\,g^{R}_{a,0}({\bar	t}_1,\bar{t}_2)\Big]
\Big[g^<_{a,0}(\bar{t}_2,t')\,g^<_{b,0}(\bar{t}_2,t')
\Big]
\nonumber\\
&&\qquad +\Big[ig^{R}_{a,0}(t,\bar{t}_2)
\,g^{R}_{b,0}(t,\bar{t}_2)\Big]\,
\Big[ig^{R}_{a,0}({\bar	t}_2,{\bar t}_1)\,
iV_{ab}^{S\,<}({\bar t}_1,{\bar	t}_2)
\,g^{R}_{b,0}({\bar	t}_2,\bar{t}_1)\Big]
\Big[g^<_{a,0}(\bar{t}_1,t')\,g^<_{b,0}(\bar{t}_1,t')\Big]\bigg\}\,.
\nonumber
\end{eqnarray}
Renaming the integration variables we arrive at
\begin{eqnarray}
\int d{\bar	t}d{\tilde t}\,&&
\Big[ig^{R}_{a,0}(t,\bar{t})\,g^{R}_{b,0}(t,\bar{t})\Big]\,
\bigg\{	V_{ab}\delta({\bar t}-{\tilde t})
+\Big[ig^{R}_{b,0}({\bar t},{\tilde	t})\,
iV_{ab}^{S\,>}({\bar t},{\tilde	t})\,
g^{R}_{a,0}({\bar t},\tilde{t})\Big]\\
&&\qquad\quad
+\Big[ig^{R}_{a,0}({\bar t},{\tilde	t})\,
iV_{ab}^{S\,<}({\tilde t},{\bar	t})\,g^{R}_{b,0}({\bar
t},{\tilde t})\Big]\bigg\}
\Big[g^<_{a,0}(\tilde{t},t')\,g^<_{b,0}(\tilde{t},t')
\Big]\,.
\nonumber
\end{eqnarray}
Thus this term belongs to the anticipated structure
\begin{eqnarray}
\int d{\bar	t}d{\tilde t}\,{\cal G}^R_{ab}(t,\bar{t})\,
\Sigma^R_{ab}({\bar	t},{\tilde t})\,{\cal G}^<_{ab}(\tilde{t},t')
\,.
\end{eqnarray}
Making the same	analysis for all term of $I_1^{(3)}$, one gets
\begin{eqnarray}
I_1^{(3)}(t,t')&=&\int d{\bar t}d{\tilde t}\,
ig^{R}_{a,0}(t,\bar{t})\,g^{R}_{b,0}(t,\bar{t})\,
\bigg\{	V_{ab}\delta({\bar t}-{\tilde t})
+ig^{R}_{b,0}({\bar	t},{\tilde t})\,
iV_{ab}^{S\,>}({\bar t},{\tilde	t})\,
g^{R}_{a,0}({\bar t},\tilde{t})\\
&&\,\,\qquad\quad
+ig^{<}_{b,0}({\bar	t},{\tilde t})\,
iV_{ab}^{S\,R}({\bar t},{\tilde	t})\,
g^{R}_{a,0}({\bar t},\tilde{t})
+ig^{R}_{a,0}({\bar	t},{\tilde t})\,
iV_{ab}^{S\,<}({\tilde t},{\bar	t})\,g^{R}_{b,0}({\bar
t},{\tilde t})\nonumber\\
&&\,\,\qquad\quad
+ig^{<}_{a,0}({\bar	t},{\tilde t})\,
iV_{ab}^{S\,A}({\tilde t},{\bar	t})\,g^{R}_{b,0}({\bar
t},{\tilde t})\bigg\}
g^<_{a,0}(\tilde{t},t')\,g^<_{b,0}(\tilde{t},t')\nonumber\\
&+&\int	d{\bar t}d{\tilde t}\,
ig^{R}_{a,0}(t,\bar{t})\,g^{R}_{b,0}(t,\bar{t})\,
\bigg\{
g^{<}_{b,0}({\bar t},{\tilde t})\,
iV_{ab}^{S\,<}({\bar t},{\tilde	t})\,
g^{<}_{a,0}({\bar t},\tilde{t})\nonumber\\
&&\,\,\qquad\quad
+g^{<}_{a,0}({\bar t},{\tilde t})\,
iV_{ab}^{S\,>}({\tilde t},{\bar	t})\,g^{<}_{b,0}({\bar
t},{\tilde t})\bigg\}
(-i)g^A_{a,0}(\tilde{t},t')\,g^A_{b,0}(\tilde{t},t')\nonumber\\
&+&\int	d{\bar t}d{\tilde t}\,
g^{<}_{a,0}(t,\bar{t})\,g^{<}_{b,0}(t,\bar{t})\,
\bigg\{	V_{ab}\delta({\bar t}-{\tilde t})
+(-i)g^{A}_{a,0}({\bar t},{\tilde t})\,
iV_{ab}^{S\,<}({\tilde t},{\bar	t})\,
g^{A}_{b,0}({\bar t},\tilde{t})\nonumber\\
&&\,\,\qquad\quad
+(-i)g^{A}_{a,0}({\bar t},{\tilde t})\,
V_{ab}^{S\,R}({\tilde t},{\bar t})\,
ig^{<}_{b,0}({\bar t},\tilde{t})
+(-i)g^{A}_{b,0}({\bar t},{\tilde t})\,
iV_{ab}^{S\,>}({\bar t},{\tilde	t})\,g^{A}_{a,0}({\bar
t},{\tilde t})\nonumber\\
&&\,\,\qquad\quad
+(-i)g^{A}_{b,0}({\bar t},{\tilde t})\,
V_{ab}^{S\,A}({\bar	t},{\tilde t})\,ig^{<}_{a,0}({\bar
t},{\tilde t})\bigg\}
(-i)g^A_{a,0}(\tilde{t},t')\,g^A_{b,0}(\tilde{t},t')\,.
\nonumber
\end{eqnarray}
A comparison with the structure	(\ref{struct1})	gives 4	additional
terms for $\Sigma^R_{ab}$ ($\Sigma^A_{ab}$), and two further terms of
$\sigma^<_{ab}$. Alltogether we	get	the	expressions	(\ref{retard-s})
for	$\Sigma_{ab}^R$	and	(\ref{sigma-kl}) for $\sigma_{ab}^<$.
%%%%%%%%%%%%%%%%%%%%%%%%%%%%%%%%%%%%%%%%%%%%%%%%%%%%%%%%%%%%%%%%%%%%%%
\subsection{Second--order contributions}\label{sec-or}
%%%%%%%%%%%%%%%%%%%%%%%%%%%%%%%%%%%%%%%%%%%%%%%%%%%%%%%%%%%%%%%%%%%%%%%
According to Eqs.~(\ref{struct1}) and (\ref{struct2})
the	second-order terms should lead to
diagrams with two self-energy insertions of	first order	with respect
to $V^S$ (reducible	diagrams)
as well	as to diagrams with	one	self-energy	insertion which
is of second order.	We will	demonstrate	this here for one typical
term. The analysis of the ladder term with two rungs (cf.
Fig.~\ref{diagram})	leads, among many other	terms, to the following
contribution ($t_1,t_2,{\bar t}_1,{\bar	t_2}$ are
integration	variables)
\begin{eqnarray}\label{I2}
I_2&=&\int\,g^{R}_{a,0}(t,t_1)\,g^{R}_{b,0}(t,t_2)\,
iV_{ab}^{S}(t_1,t_2)\,
g^{R}_{a,0}(t_1,{\bar t}_1)\,g^{R}_{b,0}(t_2,{\bar
t}_2)\\
&&\times \,iV_{ab}^{S}({\bar t}_1,{\bar
t}_2)\,g^{<}_{a,0}({\bar t}_1,t')\,g^<_{b,0}({\bar
t}_2,t')\,.
\nonumber
\end{eqnarray}

The	procedure to achieve a two-time	structure is similar to	that in
the	foregoing subsection. According	to (\ref{vs}) each causal function
$V^S$ consists of 3	terms what leads to	9 terms	in Eq.~(\ref{I2}). All
contributions containing at	least one time-diagonal	part are easily
shown to be	reducible. Therefore we	concentrate	on the others
\begin{eqnarray}\label{I2'}
I_2'&=&\int\,g^{R}_{a,0}(t,t_1)\,g^{R}_{b,0}(t,t_2)\,
i[\Theta(t_1-t_2) V_{ab}^{>}(t_1,t_2)+\Theta(t_2-t_1)
V_{ab}^{<}(t_1,t_2)]\,
g^{R}_{a,0}(t_1,{\bar t}_1)\,g^{R}_{b,0}(t_2,{\bar
t}_2)\,
\nonumber\\
&&\times i[\Theta({\bar	t}_1-{\bar t}_2)
V_{ab}^{>}({\bar t}_1,{\bar	t}_2)
+\Theta({\bar t}_2-{\bar t}_1)
V_{ab}^{<}({\bar t}_1,{\bar	t}_2)]
\,g^{<}_{a,0}({\bar	t}_1,t')\,g^<_{b,0}({\bar t}_2,t')\,.
\end{eqnarray}
These terms	should fit into	the	following structure
\begin{eqnarray}
&&\int d{\bar t}d{\tilde t}\,{\cal G}^R_{ab}(t,\bar{t})\,
\Sigma^R_{ab\,(2)}({\bar t},{\tilde	t})\,{\cal G}^<_{ab}(\tilde{t},t')
\\
&&\!\!\!+\int d{\bar t}d{\bar t}_1d{\tilde t}_1d{\tilde	t} \,
{\cal G}^R_{ab}(t,\bar{t})\,
\Sigma^R_{ab\,(1)}({\bar t},{\bar t}_1)\,
{\cal G}^R_{ab}(\bar{t}_1,\tilde{t}_1)\,
\Sigma^R_{ab\,(1)}(\tilde{t}_1,{\tilde
t})\,{\cal G}^<_{ab}(\tilde{t},t')\nonumber
\,,
\end{eqnarray}
where $\Sigma^R_{ab\,(2)}$ denotes the two-particle	self-energy	in
second order, and $\Sigma^R_{ab\,(1)}$ are the first-order quantities
identified in the foregoing	subsection.

\begin{figure}[h]
\centerline{\psfig{figure=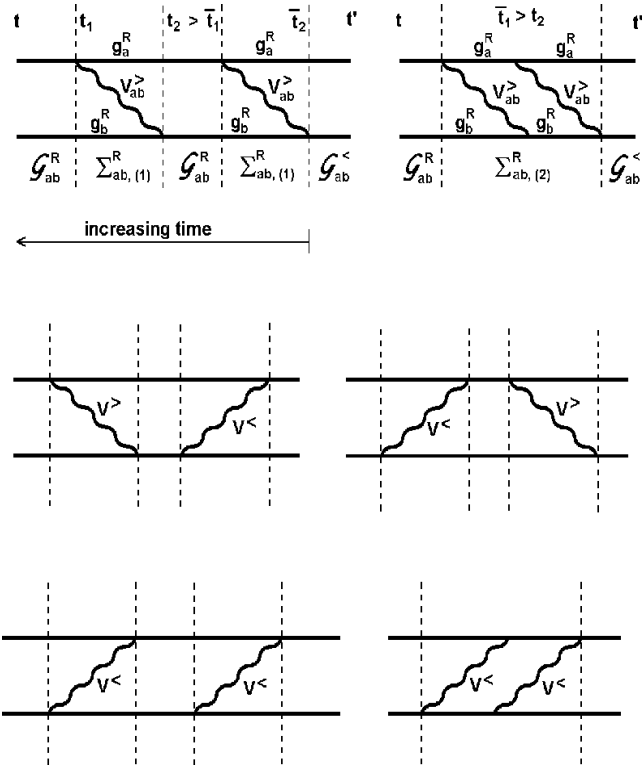,width=8.5cm,angle=0}}
\vspace*{0.5cm}
\caption[]{
Evaluation of the ladder terms with	two	rungs. The first, the third,
the	fourth,	and	the	fifth term are reducible, i.e.,	there are two
successive self-energy insertions of first order. The second and the
last term, however,	are	not	reducible: they	are	two-particle
self-energies of second	order in $V^S$.
}\label{ordnung2}
\end{figure}

Analyzing the expressions in Eq.~(\ref{I2'}), we find that the
``mixed'' terms	(with one $V^<$	and	one	$V^>$) are reducible. The
terms
containing two functions $V^>$ (or two functions $V^<$)	lead to	a
reducible term for
$t_2>{\bar t}_1$ ($t_1>{\bar t}_2$)	and	to an irreducible part for
$t_2<{\bar t}_1$ ($t_1<{\bar t}_2$). This is shown in
Fig.~\ref{ordnung2}	in form	of diagrams. The reducible terms contain
two-particle self-energy insertions	of first order with	respect	to
$V^S$. The second-order	terms contributing to $\Sigma^R_{ab\,(2)}$ are
given by
\begin{eqnarray}
\Sigma^R_{ab\,(2)}(t,t')
&=&
\int d t_1 d t_2\, \Theta(t_1-t_2)\, g^R_{b,0}(t,t_2)\,V^>_{ab}(t,t_2)
\,g^R_{a,0}(t,t_1)\,g^R_{b,0}(t_2,t')\,V^>_{ab}(t_1,t')
\,g^R_{a,0}(t_1,t')
\nonumber\\
&&+[11'a\longleftrightarrow	22'b]
\,.
\end{eqnarray}
This term can be shown to be a vertex correction to	the	two-particle
vertex.

The	other second-order diagrams	in Fig.~\ref{diagram} can be discussed
in a similar way. The terms	with two single-particle self-energy
insertions for the same	particle are reducible in any case.	For	the
other two types	of diagrams, there are reducible as	well as
irreducible	parts. This	is shown in	Fig.~\ref{classif}.	The	first
diagram	in each	row	is a reducible one.	The	second diagram is not
reducible and it corresponds to	a vertex correction	term. The third
diagram	is not reducible as	well, but is the first self-energy
correction to the diagrams of the two-particle self-energy of first
order (cf. Fig.	\ref{sigma-vs}).

Because	we started from	a ladder equation we do	not	find all possible
second-order terms contributing	to the two-particle	self-energy.
Therefore we will restrict ourselves to	the	self-energy	in first order
with respect to	the	dynamically	screened interaction.
%%%%%%%%%%%%%%%%%%%%%%%%%%%%%%%%%%%%%%%%%%%%%%%%%%%%%%%%%%%%%%%%%%%%%%
\subsection{Analysis for the other correlation functions}
\label{andere}
%%%%%%%%%%%%%%%%%%%%%%%%%%%%%%%%%%%%%%%%%%%%%%%%%%%%%%%%%%%%%%%%%%%%%%
The	analysis for the functions
$g_{ab}^{+--+}$, $g_{ab}^{-++-}$, and $g_{ab}^{--++}$ can be made in
the	same way as	above. It is scetched here for $g_{ab}^{+--+}$.
Consider the first rung	diagram. Evaluation	on the Keldysh contour
gives in analogy to	Eq.~(\ref{I3-kleiner})
\begin{eqnarray}\label{I3+--+}
I_1^{(3)}&=&g^{R}_{a,0}\,g^{R}_{b,0}\,iV_{ab}^{S}\,g^{<}_{a,0}
\,g^>_{b,0}
+g^{R}_{a,0}\,g^{<}_{b,0}\,iV_{ab}^{S\,R}\,g^{<}_{a,0}\,g^>_{b,0}
+g^{<}_{a,0}\,g^{R}_{b,0}\,iV_{ab}^{S\,A}\,g^{<}_{a,0}\,g^>_{b,0}
\\[2ex]
&&+g^{R}_{a,0}\,g^{>}_{b,0}\,iV_{ab}^{S}\,g^{<}_{a,0}\,g^{A}_{b,0}
+g^{<}_{a,0}\,g^{R}_{b,0}\,i{\overline V}_{ab}^{S}\,g^{A}_{a,0}
\,g^{>}_{b,0}
\nonumber
\\[2ex]
&&+g^{<}_{a,0}\,g^{>}_{b,0}\,iV_{ab}^{S\,R}\,g^{A}_{a,0}\,g^{<}_{b,0}
+g^{<}_{a,0}\,g^{>}_{b,0}\,iV_{ab}^{S\,A}\,g^{<}_{a,0}\,g^{A}_{b,0}
+g^{<}_{a,0}\,g^{>}_{b,0}\,i{\bar V}_{ab}^{S}\,g^{A}_{a,0}
\,g^{A}_{b,0}
\,.\nonumber
\end{eqnarray}
Again the two-time structure can be	achieved, and the structure	is
(cf. Eq. \ref{struct1})
\begin{eqnarray}
g^{+--+\,(1)}_{ab}&=&{\cal G}^R_{ab}\Big[V_{ab}+\Sigma^R_{ab}\Big]
{\cal G}_{ab}^{+--+}
+{\cal G}^R_{ab}\,\sigma^{+--+}_{ab}\,{\cal	G}_{ab}^A
+{\cal G}^{+--+}_{ab}\Big[V_{ab}+\Sigma^A_{ab}\Big]{\cal G}_{ab}^A
\end{eqnarray}
with ${\cal	G}^{+--+}_{ab}=g_{a,0}^<\,g_{b,0}^>$. The fourth and the
fivth term in (\ref{I3+--+}) are contributing to $\sigma^{+--+}_{ab}$
the	following:
$g_{b,0}^>V_{ab}^{S}g_{a,0}^<+g_{a,0}^<{\overline
V}_{ab}^{S}g_{b,0}^>$. Together	with the respective	contributions from
the	diagrams involving single-particle self-energies,
$\sigma^{+--+}_{ab}$ is	given by Eq.~(\ref{sigma+--+}) then.
%%%%%%%%%%%%%%%%%%%%%%%%%%%%%%%%%%%%%%%%%%%%%%%%%%%%%%%%%%%%%%%%%%%
\section{Kramers-Kronig	relation for the two-particle
self-energy}\label{app-hermite}
%%%%%%%%%%%%%%%%%%%%%%%%%%%%%%%%%%%%%%%%%%%%%%%%%%%%%%%%%%%%%%%%%%%
Hermitean and antihermitean	parts of the self-energy can be
constructed	in the usual way by
\begin{eqnarray}
\Sigma^{\cal H}_{ab}(121'2',\omega T)&=&
\frac{1}{2}\Big[\Sigma^{R}_{ab}(121'2',\omega T)
+\Sigma^{R}_{ab}(1'2'12,\omega T)^\ast]\,,\\
\Sigma^{\cal A}_{ab}(121'2',\omega T)&=&
\frac{1}{2}\Big[\Sigma^{R}_{ab}(121'2',\omega T)
-\Sigma^{R}_{ab}(1'2'12,\omega T)^\ast]\,.	 \nonumber
\end{eqnarray}
Due	to $\Sigma^{R}_{ab}(121'2',\omega
T)=\Sigma^{A}_{ab}(1'2'12,\omega T)^\ast$ we have
\begin{eqnarray}
\Sigma^{\cal H}_{ab}(121'2',\omega T)
&=&\frac{1}{2}\Big[\Sigma^{R}_{ab}(121'2',\omega T)
+\Sigma^{A}_{ab}(121'2',\omega T) \Big]\,,\\
\Sigma^{\cal A}_{ab}(121'2',\omega T)
&=&\frac{1}{2}\Big[\Sigma^{R}_{ab}(121'2',\omega T)
-\Sigma^{A}_{ab}(121'2',\omega T) \Big]	\nonumber\,.
\end{eqnarray}

Because	of the causality there exist the following Kramers-Kronig
relations (see,	e.g., \cite{melrose})
\begin{eqnarray}\label{kramers-kronig}
\Sigma^{\cal H}_{ab}(\omega	T)-\Sigma^{{\cal H}\,0}_{ab}(T)
&=&i{\rm P}\int_{-\infty}^\infty\frac{d\omega '}{\pi}
\frac{\Sigma^{\cal A}_{ab}(\omega '	T)
-\Sigma^{{\cal A}\,0}_{ab}(T)}
{\omega	-\omega	'}\\
\mbox{and}\nonumber\\
\Sigma^{\cal A}_{ab}(\omega	T)-\Sigma^{{\cal A}\,0}_{ab}(T)
&=&i{\rm P}\int_{-\infty}^\infty\frac{d\omega '}{\pi}
\frac{\Sigma^{\cal H}_{ab}(\omega '	T)
-\Sigma^{{\cal H}\,0}_{ab}(T)}
{\omega	-\omega	'}\,,
\end{eqnarray}
where ${\rm	P}$	denotes	the	principle value.
The	static parts,
\begin{eqnarray}
\Sigma^{{\cal H}\,0}_{ab}(T)
&=&\Sigma_a^{HF}+\Sigma_b^{HF}+\frac{1}{2}(N_{ab}-1)V_{ab}
+\frac{1}{2}V_{ab}(N_{ab}-1)\,,\\
\Sigma^{{\cal A}\,0}_{ab}(T)
&=&\frac{1}{2}(N_{ab}-1)V_{ab}
-\frac{1}{2}V_{ab}(N_{ab}-1)\,,
\end{eqnarray}
have to	be substracted in order	to ensure a	proper
behaviour in the frequency integrals.

In the approximation which is used in this paper
both contributions to the two-particle self-energy,	$\Delta_{ab}$ and
$V_{ab}^{\rm eff}$,	can	be split into hermitean	and	antihermitean
parts. For $\Delta_{ab}$ there follows
\begin{eqnarray}\label{delta-hermite}
&&\Delta^{\cal H}_{ab}({\bf	p}_1 {\bf p}_2,{\bf	p}_1'{\bf p}_2',\Omega)
=(2\pi)^6\,	\delta_{{\bf p}_1,{\bf p}_1'}\,
\delta_{{\bf p}_2,{\bf p}_2'}
\int\frac{d^3 q}{(2\pi)^3}\,
{\rm P}\int\limits_{-\infty}^\infty	\frac{d\omega_1}{\pi}\,
\Big[
-{\rm Im}\,{\varepsilon^{R}}^{-1}({\bf q},\omega_1)
\Big]\\
&&\qquad\times\bigg\{
V_{aa}({\bf	q})\frac{1\pm f_{a,{\bf	p}_1+{\bf q}}+n_B(\omega_1)}
{\Omega-\omega_1-\epsilon_{a,{\bf p}_1+{\bf	q}}
-\epsilon_{b,{\bf p}_2}}
+(a\longleftrightarrow b\,,\,1\longleftrightarrow2\,)
\bigg\}
\nonumber
 \,,
\end{eqnarray}
and
\begin{eqnarray}\label{delta-antihermite}
&&\Delta^{\cal A}_{ab}({\bf	p}_1 {\bf p}_2,{\bf	p}_1'{\bf p}_2',\Omega)
=i\,(2\pi)^6\, \delta_{{\bf	p}_1,{\bf p}_1'}\,
\delta_{{\bf p}_2,{\bf p}_2'}
\int\frac{d^3 q}{(2\pi)^3}\,
{\rm Im}\,
{\varepsilon^{R}}^{-1}
\Big(
{\bf q},\Omega-\epsilon_{a,{\bf	p}_1+{\bf q}}-\epsilon_{b,{\bf p}_2}
\Big)
\\
&&\qquad\times\bigg\{
V_{aa}({\bf	q})
\Big[
1\pm f_{a,{\bf p}_1+{\bf q}}
+n_B
\Big(
\Omega-\epsilon_{a,{\bf	p}_1+{\bf q}}-\epsilon_{b,{\bf p}_2}
\Big)\Big]
+(a\longleftrightarrow b\,,\,1\longleftrightarrow2\,)
\bigg\}
\nonumber
 \,.
\end{eqnarray}
Because	the	matrix elements	of $\Delta$	are	diagonal for homogeneous
systems, the hermitean part	is a real quantity whereas the
antihermitean part is imaginary.

Hermitean and antihermitean	parts of $V_{ab}^{\rm eff}$	are	given by
\begin{eqnarray}\label{veff-hermite}
&&V_{ab}^{\rm eff\,\cal	H}({\bf	p}_1 {\bf p}_2,{\bf	p}_1'{\bf
p}_2',\Omega)
= (2\pi)^3\,
\delta_{{\bf p}_1+{\bf p}_2,{\bf p}_1'+{\bf	p}_2'}V_{ab}({\bf p}_1-{\bf	p}_1')
\\
&&\quad\times
\Bigg\{\Bigg[
{\rm P}\int\limits_{-\infty}^\infty	\frac{d\omega_1}{\pi}\,
\Big[-{\rm Im}\,
{\varepsilon^{R}}^{-1}
({\bf p}_1-{\bf	p}_1',\omega_1)\Big]\,
\frac{1+n_B(\omega_1)\pm \frac{1}{2} f_{a,{\bf p}_1}
\pm	\frac{1}{2}	f_{b,{\bf p}_2'}}
{\Omega-\omega_1-\epsilon_{a,{\bf p}_1}-\epsilon_{b,{\bf p}_2'}}
\nonumber \\
&&\qquad +i	\,
{\rm Im}\,{\varepsilon^{R}}^{-1}
({\bf p}_1-{\bf	p}_1',\Omega-\epsilon_{a,{\bf p}_1}-
\epsilon_{b,{\bf p}_2'})
\frac{\pm 1}{2}\Big(f_{a,{\bf p}_1}-f_{b,{\bf p}_2'}\Big)\Bigg]
+(a\longleftrightarrow b\,,\,1\longleftrightarrow2\,)
\bigg\}
\,,	  \nonumber
\end{eqnarray}
\begin{eqnarray}\label{veff-antihermite1}
&&V_{ab}^{\rm eff\,\cal	A}({\bf	p}_1 {\bf p}_2,
{\bf p}_1'{\bf p}_2',\Omega)
= (2\pi)^3\,
\delta_{{\bf p}_1+{\bf p}_2,{\bf p}_1'+{\bf	p}_2'}
V_{ab}({\bf	p}_1-{\bf p}_1')
\\
&&\quad\times
\Bigg\{\Bigg[
{\rm P}\int\limits_{-\infty}^\infty	\frac{d\omega_1}{\pi}\,
\Big[-{\rm Im}\,
{\varepsilon^{R}}^{-1}
({\bf p}_1-{\bf	p}_1',\omega_1)\Big]\,
\frac{
\pm	\frac{1}{2}\Big(f_{a,{\bf p}_1}-f_{b,{\bf p}_2'}\Big)
}
{\Omega-\omega_1-\epsilon_{a,{\bf p}_1}-\epsilon_{b,{\bf p}_2'}}
\nonumber \\
&&\qquad +i	\,
{\rm Im}\,{\varepsilon^{R}}^{-1}
\Big({\bf p}_1-{\bf	p}_1',\Omega-\epsilon_{a,{\bf p}_1}-\epsilon_{b,{\bf
p}_2'}\Big)
\Big(1+n_B\Big(\Omega-\epsilon_{a,{\bf p}_1}-\epsilon_{b,{\bf
p}_2'}\Big)\pm \frac{1}{2} f_{a,{\bf p}_1}\pm
\frac{1}{2}	f_{b,{\bf p}_2'}\Big)\Bigg]\nonumber\\
&&\qquad +(a\longleftrightarrow	b\,,\,1\longleftrightarrow2\,)
\bigg\}
\,.	  \nonumber
\end{eqnarray}
Here, the matrix elements are complex in general. Diagonal matrix
elements in	an arbitrary representation,
$\langle n|V_{ab}^{\rm eff\cal H}|n\rangle$	and
$\langle n|V_{ab}^{\rm eff\cal A}|n\rangle$, are of	course real	and
imaginary, respectively.

It is easy to see that the Kramers-Kronig relations,
(\ref{kramers-kronig}),	are	fulfilled.
\newpage
%%%%%%%%%%%%%%%%%%%%%%%%%%%%%%%%%%%%%%%%%%%%%%%%%%%%%%%%%%%%%%%%%%%%%%%
\section{Connection	of $G_{ab}^A$ to the Matsubara Green's function}
\label{analyt}
%%%%%%%%%%%%%%%%%%%%%%%%%%%%%%%%%%%%%%%%%%%%%%%%%%%%%%%%%%%%%%%%%%%%%%%
It is the purpose of this appendix to scetch for the case of
thermodynamic equilibrium the connection to	the
Matsubara technique	for	imaginary-time Green's function. Especially,
it will	be shown that the Fourier transform	of the
advanced two-particle Green's function,	which was defined in
Eq.~\ref{advanc-G},	is closely related to the analytical continuation of the
two-frequency Matsubara	Green's	function.

Imaginary time Green's functions dependend on three	times were
considered by Rajagopal	et al. \cite{raja70,raja/cohen}	in connection
with an	Bethe-Salpeter equation	for	an
electron-positron system. In this appendix the same	notation is	used.
A function in which	the	times of the creation operators	are	equal is
given by
\begin{eqnarray}
{\tilde	G}^{ep}(t_1t_2t_3)=\frac{1}{i}\langle T_W
(\psi(t_1) \phi(t_2) B(t_3))\rangle
\end{eqnarray}
with
\begin{eqnarray}
B(t_3)=\phi^\dagger(t_3)\psi^\dagger(t_3)\,.
\end{eqnarray}
This function depending	on three imaginary times is	just the function
for	which the Bethe-Salpeter equation is ``closed''. The term
``closed'' means here that both	sides of the equation contain the same
type of	functions.

For	the	Fourier	coefficient	in Matsubara representation, Rajagopal and
Cohen \cite{raja/cohen}	found the following	important
double dispersion relation
\begin{eqnarray}
{\tilde	G}^{(ep)}(\omega_\lambda^e\omega_\nu^p)=\int_{-\infty}^\infty
\frac{d\omega_1\,d\omega_2}{(2\pi)^2}
\left[
\frac{f^{(2)}(\omega_1 \omega_2)}{\omega_\lambda^e-\omega_1}
+\frac{f^{(1)}(\omega_1	\omega_2)}{\omega_\lambda^e+\omega_\nu^p
-\omega_1-\omega_2}
\right]
\frac{1}{\omega_\nu^p-\omega_2}\,,
\end{eqnarray}
with
\begin{eqnarray}
f^{(2)}(\omega_1\omega_2)=\int_{-\infty}^\infty	d(t_1-t_3)\,d(t_2-t_3)\,
e^{-i\omega_1(t_1-t_3)-i\omega_2(t_2-t_3)}\,\frac{1}{i}
\Big\langle
\Big[\psi(t_1),[B(t_3),\phi(t_2)]_{-}\,\Big]_{+}\,
\Big\rangle
\end{eqnarray}
and
\begin{eqnarray}
f^{(1)}(\omega_1\omega_2)=\int_{-\infty}^\infty	d(t_1-t_3)\,d(t_2-t_3)\,
e^{-i\omega_1(t_1-t_3)-i\omega_2(t_2-t_3)}\,\frac{1}{i}
\Big\langle
\Big[
[\psi(t_1),\phi(t_2)]_{+}\,,B(t_3)\Big]_{-}\,
\Big\rangle	  \,.
\end{eqnarray}
The	Matsubara frequencies are $\omega_\nu^a=(2\nu+1)\pi	i/\beta	+
\mu^{(a)}$.

Assume that	the	function ${\tilde
G}^{(ep)}(\omega_\lambda^e\omega_\nu^p)$ has been determined. Then the
question arises	how	to extract from	this quantity
the	physical information on	the	two-particle problem. The usual	way	is
to carry out the sum over one frequency,
\begin{eqnarray}
\frac{1}{-i\beta}\sum_\nu{\tilde
G}^{(ep)}(\Omega_\mu^{ep}\omega_\nu^p)\,,
\quad  \Omega_\mu^{ep}=\omega_\lambda^e+\omega_\nu^p
\,,
\end{eqnarray}
and	performing afterwards the analytical continuation. The summation,
however, leads to the Matsubara	coefficients of	the	causal Green's
function depending on two imaginary	times, cf. (\ref{g-causal}).

We want	here to	proceed	on an other	way. The analytical	continuation
is done	for	the	two-frequency quantity.	This function
$G^{(ep)}(z_1,z_2)$	is considered now for $z_1=\Omega_e-i\eta$ and
$z_2=\Omega_p-i\eta$ with $\eta\rightarrow 0$. Making the transformation
$\Omega_e,\Omega_p\longrightarrow \Omega_{ep},\Omega_p$	we have
\begin{eqnarray}
{\tilde	G}^{(ep)}(\Omega_{ep}-i\eta,\Omega_p-i\eta)
&=&
\int\limits_{-\infty}^\infty\frac{d\omega_1\,d\omega_2}{(2\pi)^2}
\left[
\frac{f^{(2)}(\omega_1 \omega_2)}
{\Omega_{ep}-\Omega_p-\omega_1-i\eta}
+\frac{f^{(1)}(\omega_1	\omega_2)}
{\Omega_{ep}
-\omega_1-\omega_2-i\eta}
\right]\\
&&\times\frac{1}{\Omega_p-\omega_2-i\eta}
\nonumber
\,.
\end{eqnarray}
Now	an integration is perform with respect to $\Omega_p$ what leads	to
(aditionally the variables $\omega_1,\omega_2$ are transformed into
$\omega_{12}=\omega_1+\omega_2;	\omega_2$)
\begin{eqnarray}
{\tilde	G}^{(ep)}(\Omega_{ep}-i\eta)
&=&\int\limits_{-\infty}^\infty\frac{d\Omega_p}{2\pi}\,
\int\limits_{-\infty}^\infty\frac{d\omega_{12}\,d\omega_2}{(2\pi)^2}
\left[
\frac{f^{(2)}(\omega_{12}\,	\omega_2)}
{\Omega_{ep}-\Omega_p-\omega_{12}+\omega_2-i\eta}
+\frac{f^{(1)}(\omega_{12}\, \omega_2)}
{\Omega_{ep}-\omega_{12}-i\eta}
\right]\\
&&\times\frac{1}{\Omega_p-\omega_2-i\eta}
\nonumber\\
&=&
i \int\limits_{-\infty}^\infty\frac{d\omega_{12}}{(2\pi)}
\frac{1}{\Omega_{ep}-\omega_{12}-i\eta}
\int\limits_{-\infty}^\infty\frac{d\omega_{2}}{(2\pi)}
\left[
f^{(2)}(\omega_{12}\, \omega_2)+\frac{1}{2}f^{(1)}(\omega_{12}\, \omega_2)
\right]
\,.
\label{vanish}
\end{eqnarray}
The	integration	over $\omega_2$	can	be performed and gives the Fourier
transform of two-time correlation functions
\begin{eqnarray}
\int\frac{d\omega_2}{2\pi}f(\omega_{12}\omega_2)
&=&
\int\frac{d\omega_2}{2\pi}\,\int_{-\infty}^\infty
d(t_1-t_3)\,d(t_2-t_1)\,
e^{-i\omega_{12}(t_1-t_3)-i\omega_2(t_2-t_1)}\,f(t_1,t_2,t_3)\\
&=&
\int_{-\infty}^\infty
d(t_1-t_3)\,e^{-i\omega_{12}(t_1-t_3)}\,f(t_1,t_1,t_3)
\nonumber\\
&=:&f(\omega_{12})
\nonumber
\,.
\end{eqnarray}
The	equal-time commutator $[\psi(t_1),\phi(t_1)]_{+}$ vanishes and
therefore also the contribution	with $f^{(1)}$in (\ref{vanish}).
The	final result is
\begin{eqnarray}
{\tilde	G}^{(ep)}(\Omega_{ep}-i\eta)
&=&i\int\limits_{-\infty}^\infty\frac{d\omega_{12}}{(2\pi)}\,
\frac{f^{(2)}(\omega_{12})}
{\Omega_{ep}-\omega_{12}-i\eta}
\,,
\end{eqnarray}
with
\begin{eqnarray}
f^{(2)}(\omega_{12})=\int d(t-t')\,e^{-i\omega_{12}(t-t')}
\,\frac{1}{i}
\Big\langle
\Big[\psi(t),[\phi^\dagger(t')\psi^\dagger(t'),\phi(t)]_{-}\,\Big]_{+}\,
\Big\rangle
\,.
\end{eqnarray}

This result	has	to be compared with	the	Fourier	transform of
$G_{ab}^{A}$ defined in	(\ref{advanc-G})
\begin{eqnarray}
G_{ab}^{A}(\Omega_{ab})
&=&\int\limits_{-\infty}^\infty\frac{d\omega_{12}}{(2\pi)}\,
\frac{-i}
{\Omega_{ep}-\omega_{12}-i\eta}h_{ab}(\omega_{12})
\,,
\end{eqnarray}
with
\begin{eqnarray}
h_{ab}(\omega_{12})=\int d(t-t')\,e^{-i\omega_{12}(t-t')}
\,\frac{1}{-i}
\Big\langle
\Big[
\Psi_a({\bf	r}_1,t)
,
\Big[
\Psi_b({\bf	r}_2,t)
,
\Psi_b^\dagger({\bf	r}_2',t')
\Psi_a^\dagger({\bf	r}_1',t')
\Big]_{-} \,
\Big]_{\mp}	\,
\Big\rangle
\,.
\end{eqnarray}
This is, except	a minus	sign, the same expression.

%%%%%%%%%%%%%%%%%%%%%%%%%%%%%%%%%%%%%%%%%%%%%%%%%%%%%%%%%%%%%%%%%%%%

%%%%%%%%%%%%%%%%%%%%%%%%%%%%%%%%%%%%%%%%%%%%%%%%%%%%%%%%%%%%%%%%%%%%
%Figures
%%%%%%%%%%%%%%%%%%%%%%%%%%%%%%%%%%%%%%%%%%%%%%%%%%%%%%%%%%%%%%%%%%%%
\begin{figure}[h]
\centerline{\psfig{figure=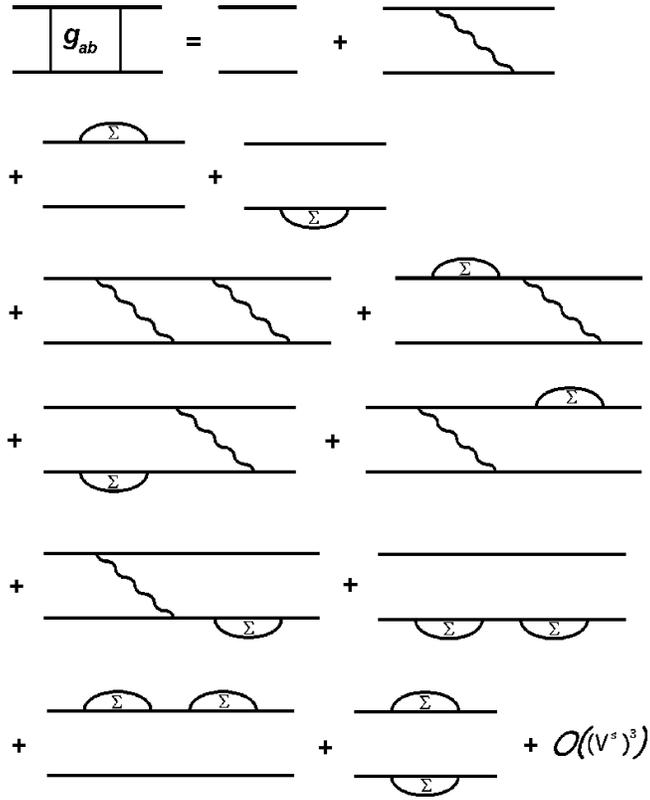,width=8.5cm,angle=0}}
\vspace*{0.5cm}
\caption[]{
Diagrammatic expansion of the dynamically screened ladder equation up
to second order.
}\label{diagram}
\end{figure}
\begin{figure}[h]
\centerline{\psfig{figure=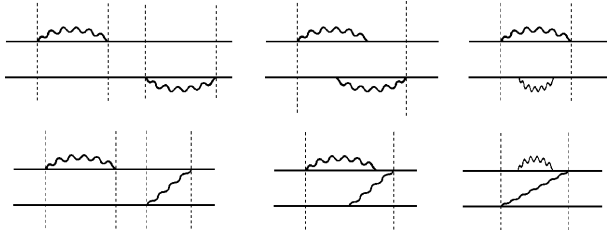,width=8.0cm,angle=0}}
\vspace*{0.5cm}
\caption[]{
Here two other types of	second-order terms are shown: in the first row
two	single-particle	self-energies are combined,	in the second row one
single-particle	self-energy	is combined	with one interaction between
the	particles. In dependence how the times do overlap, there are three
different kinds	of diagrams.
}\label{classif}
\end{figure}
%%%%%%%%%%%%%%%%%%%%%%%%%%%%%%%%%%%%%%%%%%%%%%%%%%%%%%%%%%%%%%%%%%%%
%%%%%%%%%%%%%%%%%%%%%%%%%%%%%%%%%%%%%%%%%%%%%%%%%%%%%%%%%%%%%%%%%%%%
\end{document}